\documentclass[journal=jacsat,manuscript=article]{achemso}

\usepackage{chemformula} 
\usepackage[T1]{fontenc} 
\usepackage{caption}
\usepackage{subcaption}
\usepackage[leftcaption]{sidecap}

\usepackage{soul}
\usepackage{newfloat}
\usepackage{xr}
\usepackage{hyperref}
\usepackage{amssymb}
\usepackage{tikz,xcolor,hyperref}
\usepackage{graphicx}

\makeatletter
\newcommand*{\addFileDependency}[1]{
\typeout{(#1)}
\@addtofilelist{#1}
\IfFileExists{#1}{}{\typeout{No file #1.}}
}\makeatother

\definecolor{lime}{HTML}{A6CE39}
\DeclareRobustCommand{\orcidicon}{%
	\begin{tikzpicture}
	\draw[lime, fill=lime] (0,0) 
	circle [radius=0.16] 
	node[white] {{\fontfamily{qag}\selectfont \tiny ID}};
	\draw[white, fill=white] (-0.0625,0.095) 
	circle [radius=0.007];
	\end{tikzpicture}
	\hspace{-2mm}
}

\foreach \x in {A, ..., Z}{%
	\expandafter\xdef\csname orcid\x\endcsname{\noexpand\href{https://orcid.org/\csname orcidauthor\x\endcsname}{\noexpand\orcidicon}}
}

\author{Theodoros D. Bouloumis} 
\affiliation{Light-Matter Interactions for Quantum Technologies Unit, Okinawa Institute of Science and Technology Graduate University, Onna, Okinawa 904-0495, Japan}
\email{Theodoros.Bouloumis@oist.jp}
\author{Hao Zhao}
\affiliation{Organic and Carbon Nanomaterials Unit, Okinawa Institute of Science and Technology Graduate University, Onna, Okinawa 904-0495, Japan}
\author{Nikolaos Kokkinidis}
\affiliation{Light-Matter Interactions for Quantum Technologies Unit, Okinawa Institute of Science and Technology Graduate University, Onna, Okinawa 904-0495, Japan}
\author{Yunbin Hu}
\affiliation{College of Chemistry and Chemical Engineering, Central South University, Changsha, 410083, China}
\author{Viet Giang Truong}
\affiliation{Light-Matter Interactions for Quantum Technologies Unit, Okinawa Institute of Science and Technology Graduate University, Onna, Okinawa 904-0495, Japan}
\author{Akimitsu Narita}
\affiliation{Organic and Carbon Nanomaterials Unit, Okinawa Institute of Science and Technology Graduate University, Onna, Okinawa 904-0495, Japan}
\author{S\'{i}le {Nic Chormaic}} 
\affiliation{Light-Matter Interactions for Quantum Technologies Unit, Okinawa Institute of Science and Technology Graduate University, Onna, Okinawa 904-0495, Japan}
\email{sile.nicchormaic@oist.jp}

\title[An \textsf{achemso} demo]
  {Disruptive Forces in Metamaterial Tweezers for Trapping 20 nm Nanoparticles Based on Molecular Graphene Quantum Dots}

\keywords{American Chemical Society, \LaTeX}

\begin{document}


\begin{abstract}
In recent years, plasmonic optical tweezers have been used to trap nanoparticles and study interactions with their environment. An unavoidable challenge is the plasmonic heating due to resonant excitation and the resulting temperature rise in the surrounding environment. In this work, we demonstrate trapping of custom-synthesized 20~nm nanoparticles based on molecular graphene quantum dots using metamaterial plasmonic tweezers. Superior trap stiffness values as high as 8.8~(fN/nm)/(mW/$\mu\mbox{m}^2$) were achieved with optical intensities lower than 1~mW/$\mu\mbox{m}^2$. By gradually increasing the laser intensity we identified a critical value beyond which the stiffness values dropped significantly. This value corresponded to a temperature rise of about 16$^o$C, evidently sufficient to create thermal flows and disrupt the trapping performance. We, therefore, identified a safe intensity regime for trapping nanoparticles without unwanted heat. Our platform can be used for efficient nanopositioning of fluorescent particles and quantum emitters in an array configuration, potentially acting as a single-photon source configuration.
\end{abstract}

\noindent{\bf Keywords: metamaterial plasmonic tweezers, fluorescent nanoparticles, optical trapping, thermal effects, nanographenes} 


\section{1. Introduction}
 Quantum emitters (QEs) are nanoscale artificial molecules synthesized with specific properties and a total diameter of less than 100~nm~\cite{Novotny1997Nanometric}. Some of the main types of solid-state quantum emitters are semiconductor quantum dots (QDs)~\cite{Kagan2000QDs, Reece2014ColloidalQD, Thompson2021QuantumInformation}, nanodiamonds with nitrogen vacancy (NV) centers~\cite{AharonovichDiamond2011, Gibson2019Nanodiamonds}, graphene quantum dots (GQDs)~\cite{JIN2015439, YiboGQD2019}, fluorescent organic molecules~\cite{Novotny1997Nanometric, Toninenlli2021OrganMol, Gaither2023OrgMolec}, and metal-halide perovskite nanoparticles~\cite{Maksym2015PerovskiteQDs}.

Advances in chemical synthesis methods have enabled tailoring of QE properties such as size, shape, internal structure, absorption, and emission spectra. Therefore, the ability to control those properties offers highly-tunable photon sources, exhibiting high quantum yield and emitting in a range of wavelengths from UV to IR. QEs have attracted interest from multiple fields such as biology, serving as luminescent probes and biomarkers~\cite{Reece2014ColloidalQD, Matos2020QDNanolabels, Hao2022D7H}, display industry, as highly-tunable LEDs for next-generation displays~\cite{Jang2010QDWhiteLED, Liu2020QDLED}, optoelectronics, for light-emitting devices~\cite{Kagan2016QDDevices, TakeQD2021, Ha2021Organic}, and, of course, they play a central role in the field of quantum technologies acting as single-photon sources~\cite{Lounis2000Molecule, Zhang:08, Mizu2012diamond, Imamoglu2015ReviewEntanglement, Thompson2021QuantumInformation, Carsten2022Singlephoton}.

Due to all the remarkable properties and promising applications of nanoparticle QEs, there was a desire to develop optical trapping techniques to facilitate studies on their optomechanical properties and interactions with the environment~\cite{Horowitz2012ElectronSR, Jauffred2014Subdiffraction}. The first demonstration of single QE trapping was reported by Jauffred et al. in 2008, achieved via conventional optical tweezers using a continuous wave (CW) laser in the infrared with a trapping power as high as 0.5~W to trap a \textit{CdSe/ZnS} QD~\cite{Liselotte2008QD}. This power was very high considering today's standards, and formidable for many applications as it could easily spoil the QD due to the strong laser irradiation and heat transfer~\cite{Li2022PhotothermalQD}. A few years later, the same group demonstrated trapping of various semiconductor QDs with lower laser intensity and simultaneous two-photon excitation~\cite{Liselotte2010TwophotonQD}. Aiming to reduce the incident intensity led to trapping of QDs with near-field forces utilizing plasmonic~\cite{Tsuboi2010QDGap, Bawendi2016Bowtie, Ndukaife2021ElecTherPlas} and dielectric structures~\cite{Erickson2012Photonic, Xu2019AlldielectricNF}. Such configurations can significantly reduce the incident intensity required for the trapping, while resulting in stronger trap stiffnesses~\cite{Bouloumis2020FromFT, Bouloumis2020FastAE}. Some works have also investigated  Kramers hopping in plasmonic nanostructures using trapped QDs, which are ideal due to their high refractive index~\cite{Gordon2013Hopping, Yoon2020HoppingOS}.

One drawback of QDs is their chemical compositions containing heavy metal elements, typically including cadmium~\cite{CdQDs2008} or lead~\cite{LeadQDs2017}. Therefore, GQDs have attracted attention as carbon-based and thus environmentally friendly alternative. However, the majority of the GQDs reported in the literature are prepared through solvothermal methods~\cite{TIAN2016204} and their chemical structures are typically undefined, prohibiting the establishment of precise structure-property relationship and fine-tuning of their optical properties~\cite{Photofluo}. To this end, molecular nanographenes, namely large polycyclic aromatic hydrocarbons, can be bottom-up synthesized through the methods of synthetic organic chemistry, providing atomically precise GQDs with structure-dependent properties~\cite{Narita2021Poly}. Moreover, such molecular GQDs can be assembled with amphiphilic polymers to prepare fluorescent polymeric nanoparticles that are water-soluble~\cite{Hao2022D7H}. 

In previous works, the type of molecular GQDs studied here was used for lysosome-targeted cancer photodynamic therapy~\cite{Hao2022D7H} due to their biocompatible nature. In this work, our interest was in trapping nanoparticles containing molecular GQDs and studying their positioning dynamics on a periodic metamaterial array for potential application as efficient single-photon sources in the field of quantum technologies. We present trapping of molecular GQD-based nanoparticles, also refer to them as D7H nanoparticles (D7H-NPs), using a twisted nanographene called double [7]helicene (D7H)~\cite{Narita2017D7H, Hao2022D7H}. The D7H-NPs had an average diameter of 20~nm, and they were trapped using metamaterial plasmonic tweezers~\cite{Kotsifaki2020FanoResonantAM, BouloumisEnabling}. Previous works have shown that metamaterial plasmonic tweezers are ideal for efficiently trapping sensitive particles that cannot handle high temperatures~\cite{KotsifakiColi2023, kotsifaki2024hybrid}. Recently, we used the same metamaterial plasmonic tweezers to investigate the optomechanical response of the trap using gold nanoparticles 20~nm in diameter~\cite{BouloumisEnabling}. That work revealed the existence of two types of plasmonic hotspots periodically arranged, resulting in different trap stiffnesses. In this work, we observed similar behavior when trapping the D7H-NPs. However, we progressively introduced errors in classifying the trapping hotspot location with increasing trapping laser intensity. We attributed this observation to the induced heating of the water solution (where the NPs were suspended) with increasing laser intensity. Therefore, we hypothesize that the measured trap stiffnesses at higher intensities were not a result of optical forces alone, but also from the induced thermal forces in the solution~\cite{Jiang2020QuantifyingTR, Kotsifaki2022Thermal}.

\section{2. Materials and Methods}

\par{\textbf{A. D7H-NPs Synthesis and Properties}}

\begin{figure}[ht!]
     \centering 
     \captionsetup{justification=raggedright,singlelinecheck=false}
     \begin{subfigure}[b]{0.75\textwidth}
         \centering
         \caption{}
         \includegraphics[width=\textwidth]{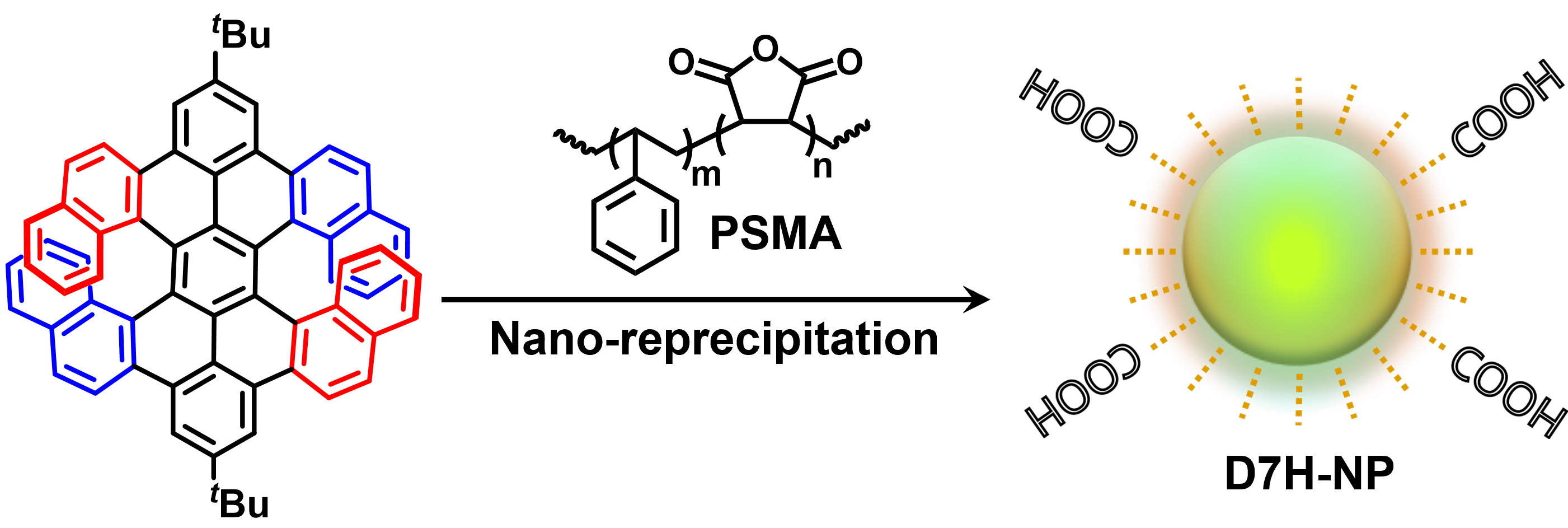}
         \label{D7HStructure}
     \end{subfigure}
    \hspace{2cm}
     \begin{subfigure}[b]{0.32\textwidth}
         \centering
         \caption{}
         \includegraphics[ width=\textwidth]{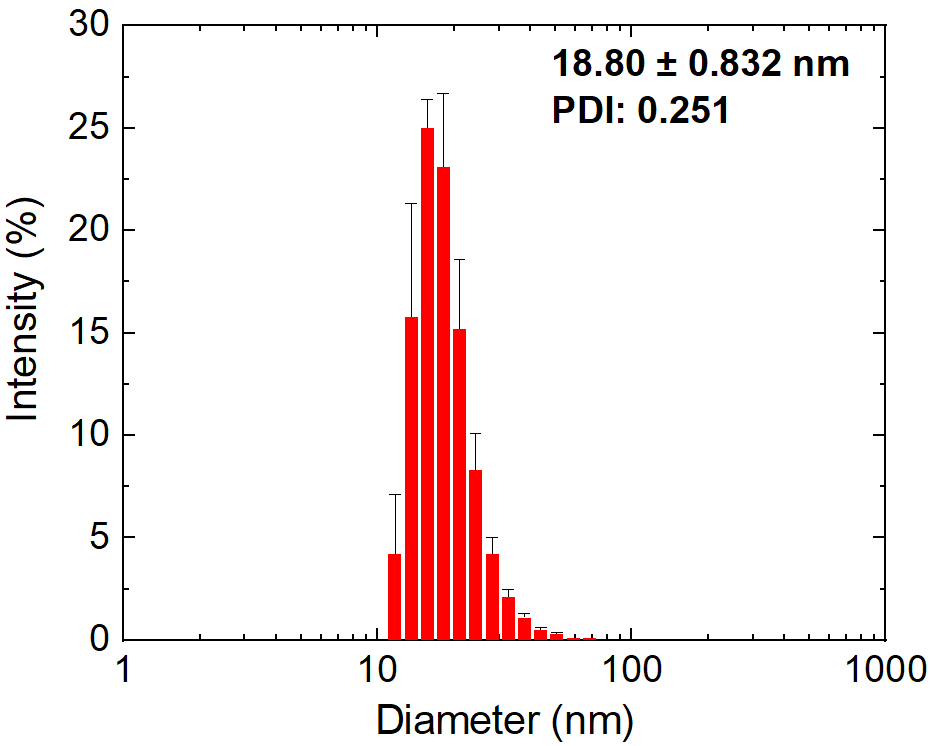}
         \label{D7HDiameter}
     \end{subfigure}
     \hfill
     \begin{subfigure}[b]{0.325\textwidth}
         \centering
         \caption{}
         \includegraphics[ width=\textwidth]{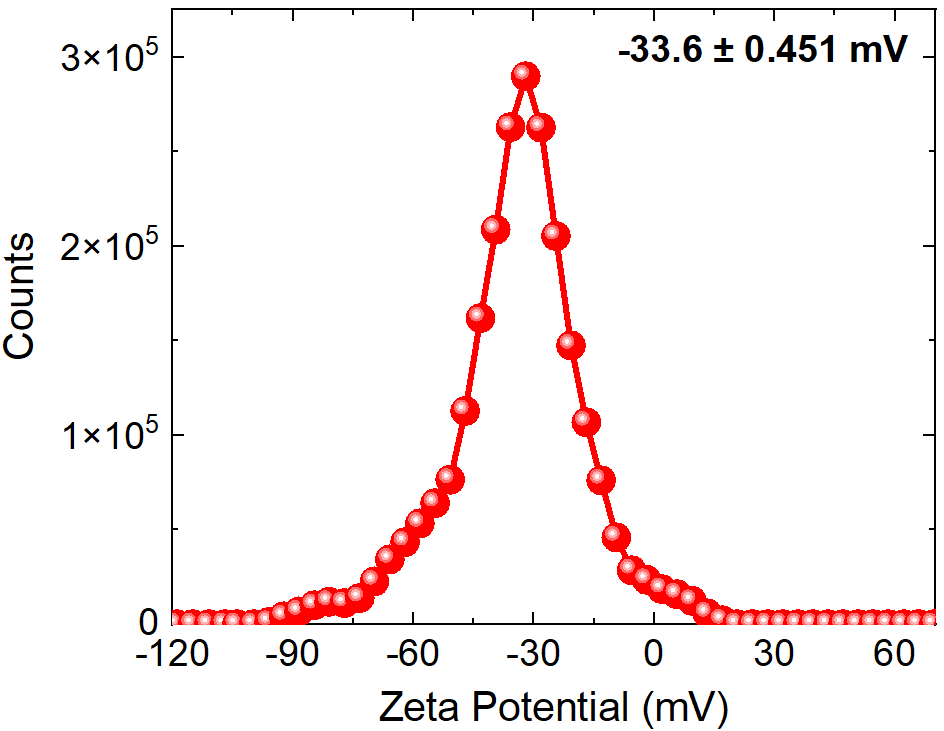}
         \label{D7Hzpotential}
     \end{subfigure}
     \hfill
     \begin{subfigure}[b]{0.325\textwidth}
         \centering
         \caption{}
         \includegraphics[ width=\textwidth]{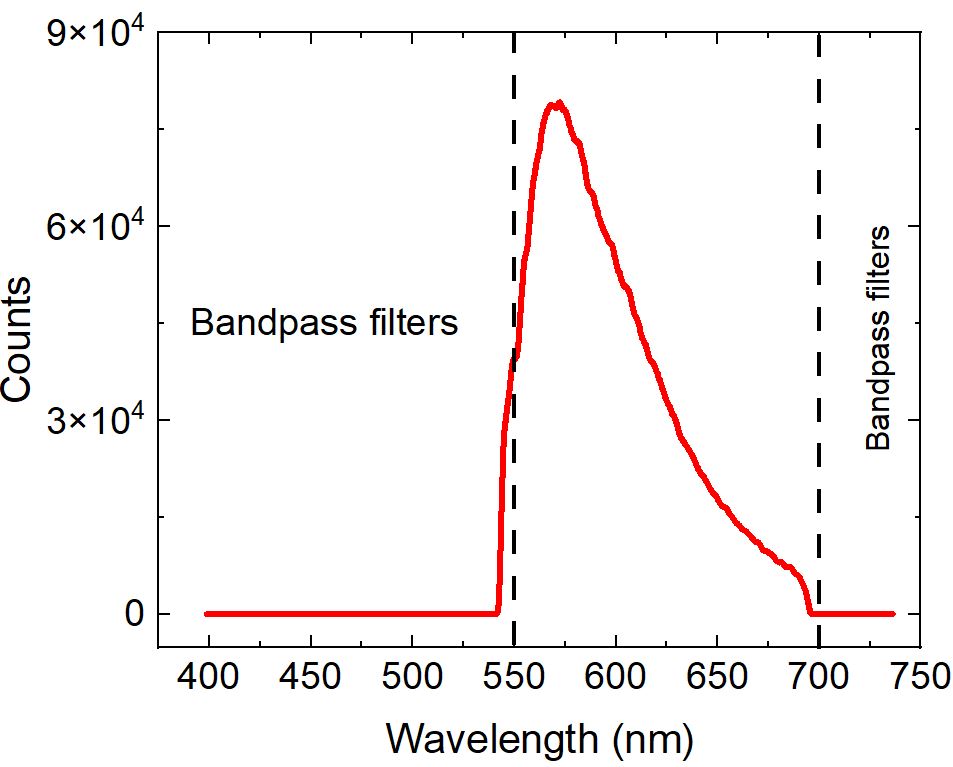}
         \label{D7HEmission}
     \end{subfigure}
    \captionsetup{justification=justified}
    \caption[Basic characteristics of the double-7-carbohelicene nanoparticles]{Basic characteristics of the D7H-NPs. (a) Chemical structure and representation of the chemical synthesis of the D7H-NPs, (b) nanoparticle size distribution, (c) zeta potential of the nanoparticles' surface, (d) emission spectrum with a peak at 575~nm.}
    \label{D7H}
\end{figure}

The details of the chemical synthesis of D7H molecules and the preparation of similar D7H-NPs were previously published~\cite{Narita2017D7H, Hao2022D7H}.  In this work, D7H-NPs having quasi-core/shell structure, with fluorescent D7H molecules inside the core and hydrophilic moieties of amphiphilic polymers, as the shell, were prepared through a classic nano-reprecipitation method~\cite{Hao2022D7H}, as shown in Figure~\ref{D7H}a. 0.25 mg of D7H molecule and 0.5~mg of PSMA (Poly(styrene-\textit{co}-maleic anhydride) purchased from Sigma Aldrich) were dissolved in 5~mL of tetrahydrofuran (THF), the obtained homogeneous solution was added into 15~mL of deionized water under sonication, and was stirred at room temperature with nitrogen bubbling for 48~h to remove THF and produce D7H-NPs. The obtained stock solution was stored in the dark at 4$^o$C. 

In Figure~\ref{D7H}, we depict the chemical structure of D7H molecule, schematic illustration of the preparation of D7H-NPs, the size distribution, zeta potential, and emission spectrum of the NPs. The size distribution was measured using the dynamic light scattering method (DLS) indicating a diameter of $18.8 \pm 0.8$~nm. This is a typical size for a quantum emitter, and can be readily trapped and manipulated using plasmonic nanostructures. The DLS measurements were performed using a Malvern Zetasizer Nano ZS 90 (Malvern, UK) instrument equipped with a He–Ne laser (633~nm, 4~mW). Measurements were carried out in a polystyrene cuvette at a 173$^o$ accumulation angle after equilibrating for 2~min at 25$^o$C. The data were processed by the instrument software (Zetasizer Nano software v3.30) to give the number mean particle size and polydispersity index value by a non-negative least squares method. The zeta potential of D7H-NPs was simultaneously measured by the  Malvern Zetasizer Nano ZS 90 instrument, yielding a mean value of $-33.6 \pm 0.5$~mV. We deliberately selected PSMA as the amphiphilic polymer to obtain D7H-NPs with negatively charged surfaces so that the NPs would not stick on the gold surface of the plasmonic nanostructures, since it is also negatively charged. However, electrostatic interactions might still have played some role in the trapping process~\cite{RodrguezSevilla2018OpticalFA}. Lastly, the emission spectrum of D7H-NPs was recorded showing a peak at around 575~nm. Emission in the visible range is useful for quantum technology applications while the NPs could be excited using blue light (490~nm). The experimental details for obtaining the spectrum are described in Section 2.3. It worth noting that the D7H-NPs are biocompatible, making them ideal for \textit{in-vivo} use in medical applications~\cite{Hao2022D7H}.
\\
\par{\textbf{B. Metamaterial Fabrication and Characterization}}

The plasmonic metamaterial array was fabricated on a 50-nm thin film of gold (PHASIS, Geneva, BioNano), using  focused ion beam milling (FIB-FEI Helios Nanolab G3UC). The gold was deposited on a borosilicate glass (SiO$_2$) substrate with a 5~nm titanium adhesion layer. The design was based on the asymmetric split-ring resonator shown in the inset of Figure~\ref{Hist_k_all}, and the desired geometrical characteristics of the structures were determined using finite element method simulations to achieve a particular resonance wavelength at around 930~nm. An array of 16$\times$16 units was fabricated with a period of 400~nm in both directions. The etched film was then treated with oxygen plasma to remove any deposited gallium residue that resulted during the FIB etching~\cite{SuKo_Ga}. Next, we performed microspectrophotometry measurements (CRAIC, 20/30 PV) to obtain the experimental transmission spectrum through the metamaterial array, and, therefore, the experimental plasmonic resonance. In our design, the measured  resonance was at 930~nm, therefore we tuned the trapping laser to this wavelength to excite a strong plasmonic field. The metamaterial design and the characterisation methods are the same as those used in our previous work, and the interested reader is directed there for a more detailed description~\cite{BouloumisEnabling}. 

\par{\textbf{C. Optical Trapping Setup}}

\begin{figure}
    \centering  \captionsetup{justification=raggedright,singlelinecheck=false}
    \begin{subfigure}[b]{\textwidth}
        \centering
        \caption{}
        \includegraphics[width=0.7\textwidth]{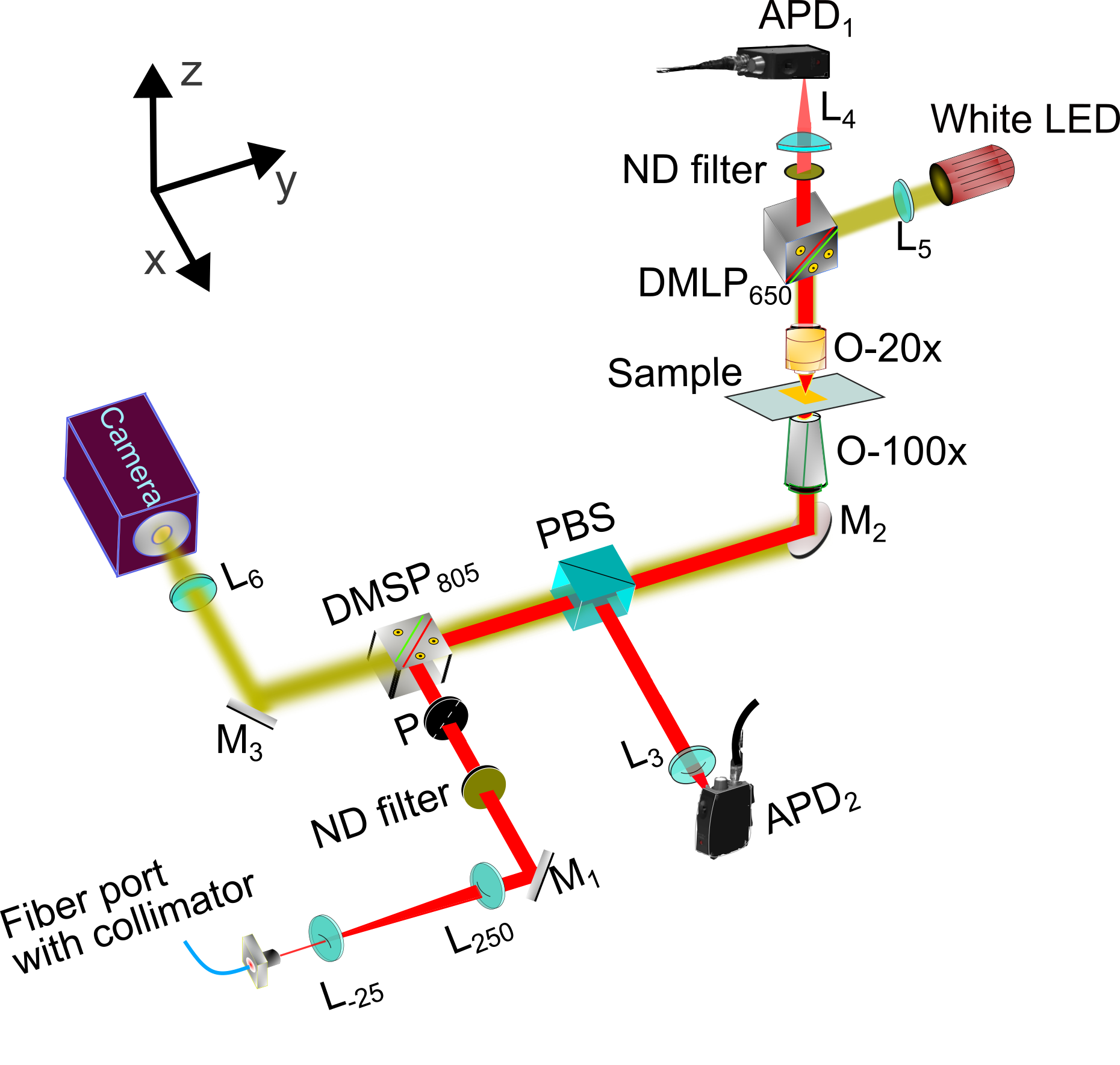}
        \label{setup}
    \end{subfigure}
    \begin{subfigure}[b]{0.6\textwidth}
        \centering
        \caption{}
        \includegraphics[width=\textwidth]{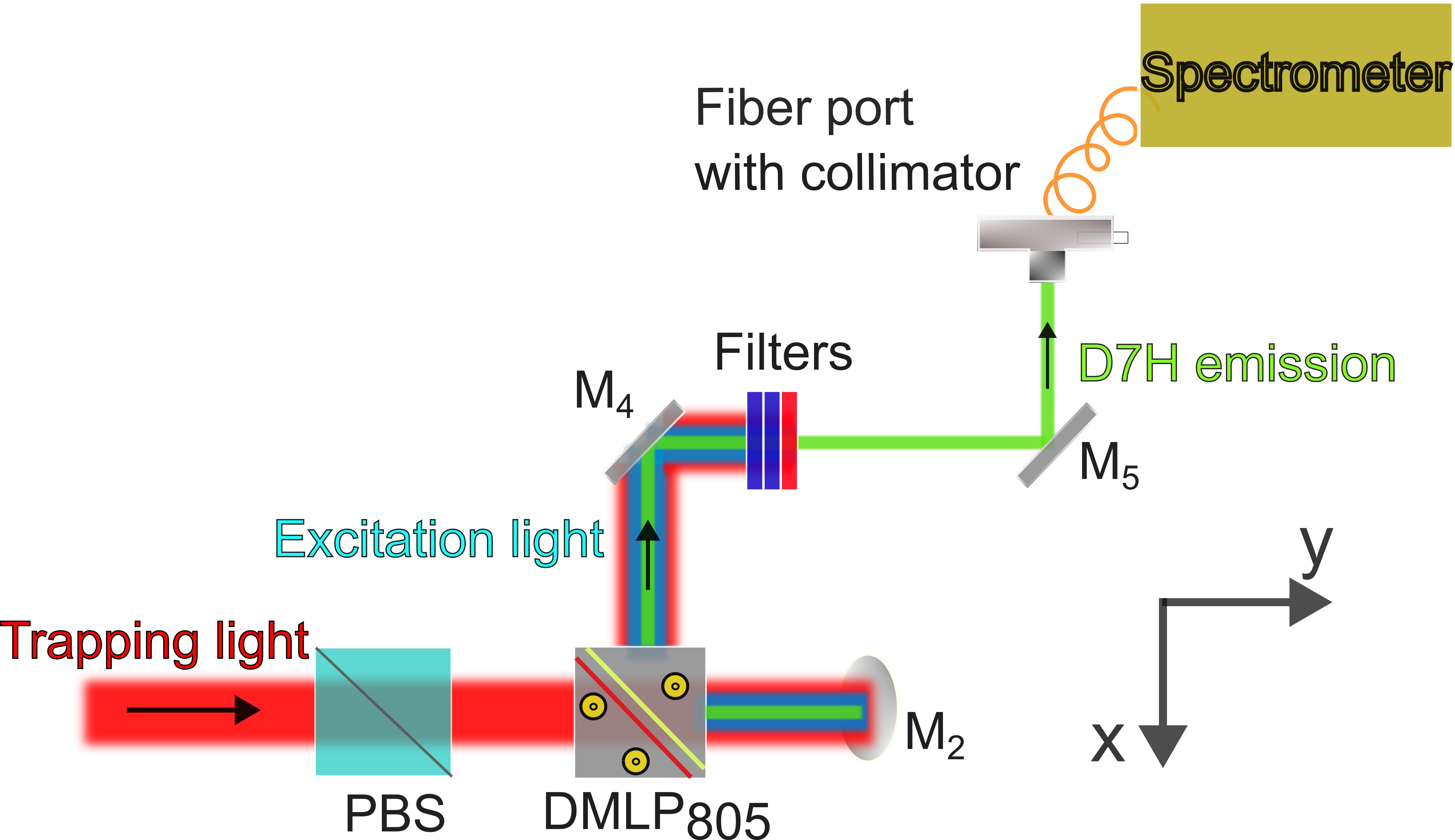}
        \label{Add_setup}
    \end{subfigure}
    \captionsetup{justification=justified}
    \caption{(a) Schematic of the basic configuration of the experimental setup used for the trapping experiments. M: mirror, L: lens, ND filter : neutral density filter, P: polarizer, DMSP and DMLP: shortpass and longpass dichroic mirrors, respectively, PBS: polarizing beamsplitter, O: objective lens, APD: avalanche photodiode. Dimensions not scaled. (b) The parts added in between the PBS and mirror M$_2$ in order to record the D7H-NPs emission spectrum.}
    \label{setup-parts}
\end{figure}

The experimental setup  for the trapping experiments is shown in Figure~\ref{setup}. Linearly polarized light from a Ti:Sapphire laser (Coherent MBR-110) tuned to 930~nm wavelength was used to excite the Fano resonance in the metamaterial array. It was coupled in the setup using a single-mode polarization-maintaining fiber. A polarizing beamsplitter (PBS) ensured the desired $y$-axis polarization direction (see inset in Figure~\ref{Hist_k_all}) while at the same time facilitated the control of incident power by rotating a linear polarizer (P). An objective lens with magnification 100$\times$ and NA 1.25 was used to focus the incident light on the metamaterial. Transmitted light through the metamaterial yielded information about trapping events, and was collected by a second objective lens with magnification 20$\times$, NA 0.46 and detected on an avalanche photodiode (APD). The APD signal was analyzed using trapping transient analysis and the trap stiffness for each trapping event was calculated. 

In order to record the emission spectrum of the D7H-NPs, the setup was modified accordingly. The white light LED shown in Figure~\ref{setup} was replaced by a laser (Toptica, iChrome, TVIS) with wavelength tuned to 488~nm for excitation of the NPs. This beam was in the opposite direction to the trapping beam. A droplet of the NPs solution was deposited on the metamaterial and allowed to evaporate. Thus, any emitted light  was due to the collective excitation of many NPs. To collect the emissions, we added some extra parts in between the PBS and mirror M$_2$ (see Figure~\ref{setup}), as shown in Figure~\ref{Add_setup}. A dichroic mirror DMLP$_{805}$ was used to guide the emitted light through a series of bandbpass and notch filters that were used to block the trapping and excitation lasers, then into a multimode fiber connected to a spectrometer (Andor, Shamrock 500i) to record the spectrum, as shown in Figure~\ref{D7HEmission}.
\\
\par{\textbf{D. Solution Preparation}}

For the trapping experiments, the D7H-NPs were dispersed in heavy water (D$_2$O) with a concentration of 0.15~$\mu$g/mL based on the total weight of D7H-NPs (D7H and amphiphilic polymer), and surfactant Tween-20 was also added with 0.1\% v/v concentration to avoid the aggregation of NPs in the solution. The final solution was shaken well and 8.5~$\mu$L were placed directly onto the gold film surface inside a seal spacer with 120~$\mu$m thickness, and enclosed on the top with a micro glass cover slip, thus forming a microwell where the NPs were suspended. The sample was then mounted in the setup with the cover slip facing upwards and the glass substrate downwards, with the trapping laser  incident from the bottom of the sample (glass substrate side). Detailed description of the experimental processes mentioned in Section~2 can be found in reference~\cite{TheoThesis}.
\\
\par{\textbf{E. Temperature Rise Simulations}}

For the temperature rise simulations, we used a commercial finite element modeling \textit{COMSOL Multiphysics 6.0} software. A 3D model was established to solve the electromagnetic (EM) and the heat transfer (HT) problems. The modeling process was governed by a set of differential equations describing the full-wave EM and HT physics, and the coupling phenomena between them. The 3D EM domain was set at 3.2$\times$3.2$\times$3.2~$\mu$m cube, consisting of an array of 6$\times$6 unit cells (see inset of Figure~\ref{Hist_k_all}) illuminated from the substrate-side by a laser with a Gaussian intensity distribution. The focal-spot radius was set at $w_0 = 0.61\lambda/\mbox{NA}$, where $\lambda$ is the wavelength of the incident laser light and $\mbox{NA} = 1.25$ the numerical aperture of the focusing objective lens. Scattering boundary  conditions were used on all outer EM domain boundaries. This allowed EM radiation to propagate out of the EM domain without reflection. Subsequently, the full-EM domain was placed inside a larger 40$\times$40$\times$40~$\mu$m cube HT domain. A prescribed ambient temperature of 293.15~K was set at the HT boundaries for solving the heat transfer problem. The input Gaussian beam was polarized parallel to the $y$-direction (inset of Figure~\ref{Hist_k_all}). Initially, the distribution of the excited electromagnetic field was simulated by numerically solving the scattering problem using the time-independent Maxwell’s vector wave equation (inset of Figure~\ref{Hist_k_all}).  Next, the heat source density was obtained via the equation $q_i(\textbf{r}) = 1/2 Re[\textbf{J}\cdot \textbf{E}^*]$, where $\textbf{J}$ is the induced-current density in each unit cell, $\textbf{E}$ is the electric field distribution, $Re$ stands for the real part of the complex number, and $*$ indicates the complex conjugate. Finally, the total heat power, $Q_{tot} = \oint q_i(\textbf{r})d^3\textbf{r}$, was obtained by taking the sum of the heat power generated at each mesh element, and served as the total power dissipation over the 3D volume of the coupled EM and HT model. The optical properties of the gold and titanium metals were defined using the Brendel-Bormann model~\cite{Rakic:98}, and the particles were treated as polystyrene with properties obtained from the COMSOL material library. The thermal properties of the glass, gold, titanium, and water were also adopted from the COMSOL material library. The model required 0.55~TB of memory, and the simulations were performed on the OIST computational cluster.

\section{3. Results and Discussion}

The trapping experiments were performed with increasing incident laser power from 1.5 - 6.5~mW on the sample, corresponding to an intensity range of approximately 0.5 - 2.1~mW/$\mu$m$^2$, since the spot size on the sample had a radius of about 1~$\mu$m. For each intensity value, the experiment took only a few minutes (less than 10~min) during which the trapping beam was blocked and released in certain time intervals, thus obtaining multiple trapping events. It is worth noting that the first trapping event occurred within just a few seconds after the trapping beam was initially switched on,  indicating fast transport and trapping of the NPs which is a unique feature of our design~\cite{Kotsifaki2020FanoResonantAM, BouloumisEnabling}. In total, more than 250 trapping events were recorded and analyzed. 

\begin{figure}
    \centering
    \includegraphics[width=0.6\textwidth]{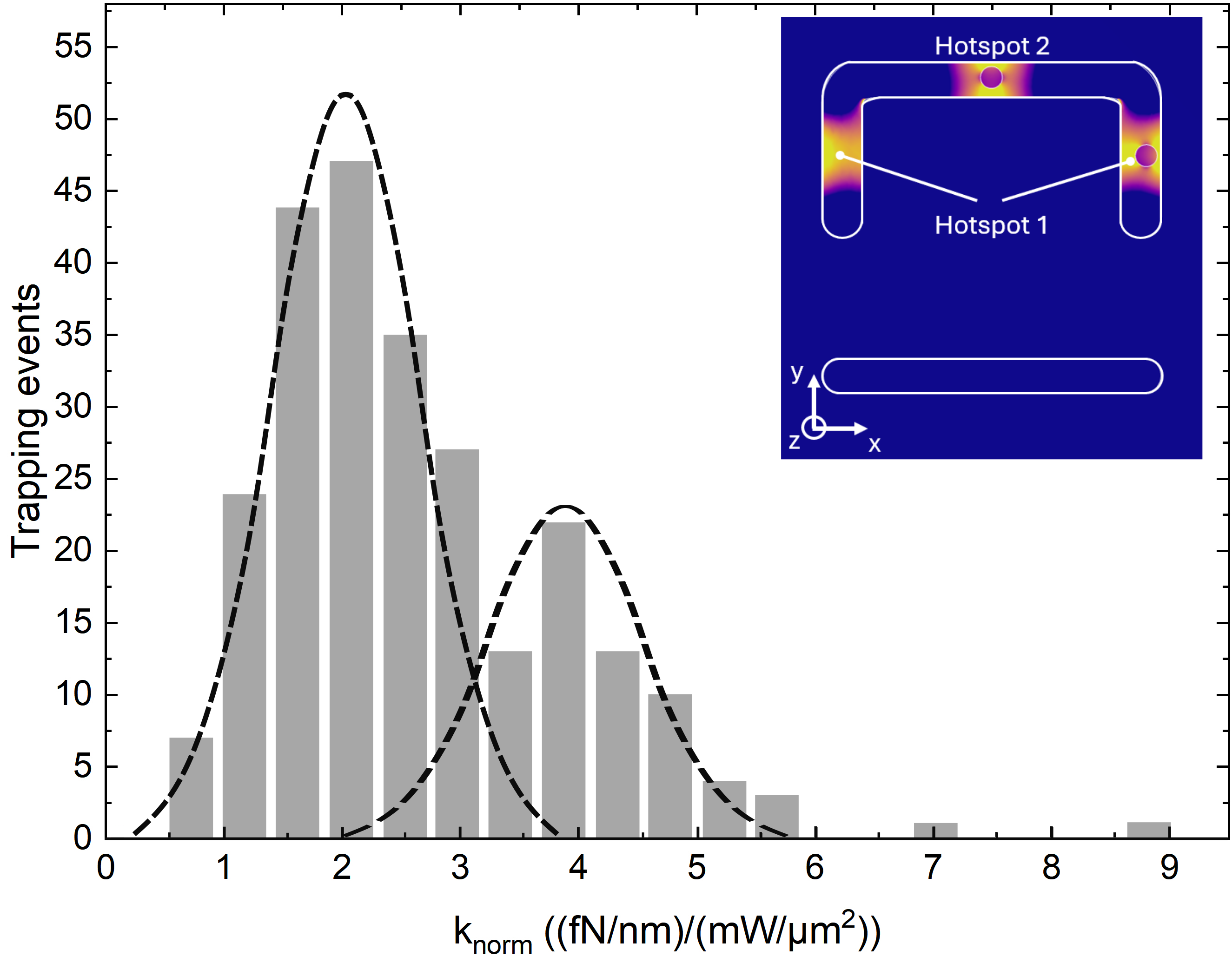}
    \caption{Histogram of the measured normalized trap stiffnesses of the trapped D7H-NPs. The trapping events included were obtained with excitation intensities ranging from 0.59 - 2.1~mW/$\mu\mbox{m}^2$ (1.56 - 6.6~mW). The dashed lines are for visual aid indicating the distribution of trap stiffness values due to the existence of two types of hotspots. Inset shows the simulated electromagnetic field on a unit cell of the metamaterial, depicting the two types of hotspots excited, with and without a trapped particle.}
    \label{Hist_k_all}
\end{figure}

In Figure~\ref{Hist_k_all}, a histogram of the normalized trap stiffness values (trap stiffness divided by the incident intensity) is plotted. The values are very high even for low incident intensities when compared with the literature~\cite{Kotsifaki2019PlasmonicOT, Bouloumis2020FromFT, Bouloumis2020FastAE}, indicating strong and efficient trapping with values as high as 8.8~(fN/nm)/(mW/$\mu\mbox{m}^2)$. The histogram seems to consist of two distributions (see dashed lines in Figure~\ref{Hist_k_all}) owing to the two different types of hotspots excited on the metamaterial array, as explained below. 

A detailed investigation of the excited hotspots and their characteristics for trapping experiments can be found in our previous work~\cite{BouloumisEnabling}. Briefly, in our metamaterial array, we can excite simultaneously two  types of hotspots (referred to as Hotspot 1 \& 2), which can be tuned by the wavelength of the excitation laser. Each unit of the metamaterial array possesses two hotspots of type 1 and one of type 2, as shown in the inset of Figure~\ref{Hist_k_all}. Therefore, we expect that, probabilistically, there will be twice as many trapping events at Hotspot 1 compared with Hotspot 2. The optical force and, as a result, the trap stiffness that a particle experiences in each of the hotspots depend on the cavity - particle coupling. Thus, parameters such as the particle's polarizability and geometrical characteristics of the nanocavity affect the trapping performance accordingly. In our work, the D7H-NPs have a refractive index close to that of a polystyrene particle and therefore, based on simulations and previous experiments, a particle trapped at Hotspot 1 experiences stronger optical forces and a higher trap stiffness compared to a particle trapped in Hotspot 2 (see reference~\cite{TheoThesis} for details). Note that for gold nanoparticles of the same size, the behavior of the hotspots is the opposite~\cite{BouloumisEnabling}.

\begin{figure}[ht]
    \centering
    \captionsetup{justification=raggedright,singlelinecheck=false}
    \begin{subfigure}[b]{0.46\textwidth}
        \centering
        \caption{}
        \includegraphics[width=\textwidth]{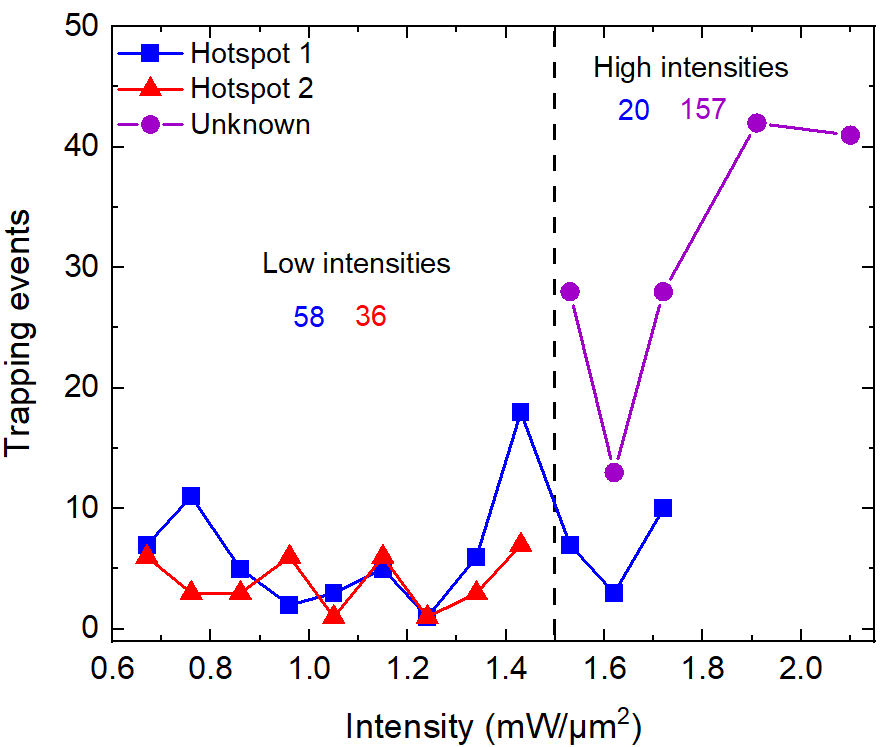}
        \label{Events_I}
    \end{subfigure}
    \hfill
    \begin{subfigure}[b]{0.47\textwidth}
        \centering
        \caption{}
        \includegraphics[width=\textwidth]{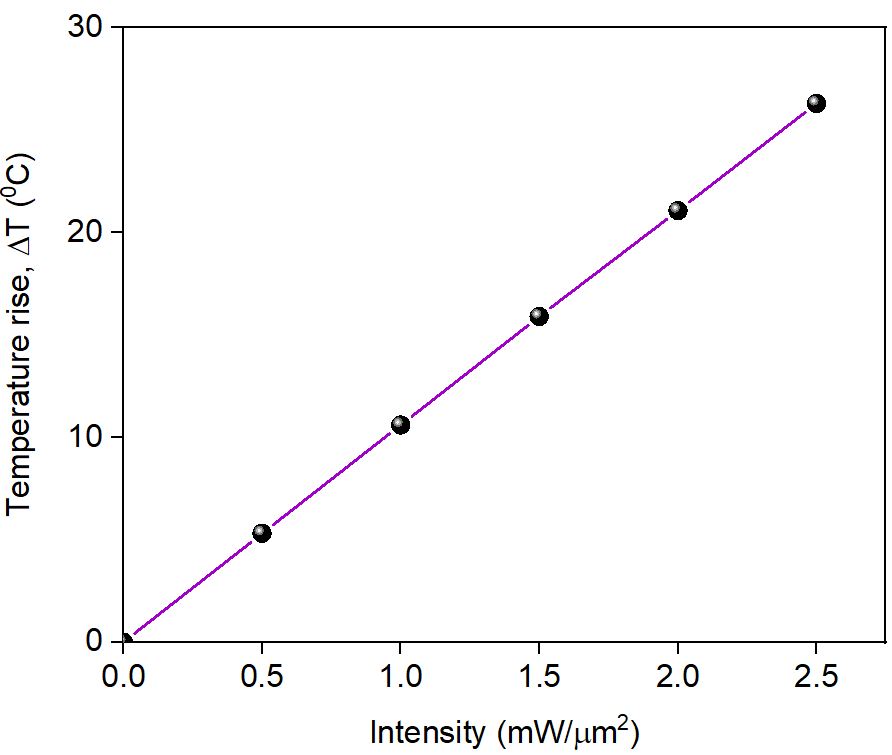}
        \label{Thermal-I}
    \end{subfigure}
    \captionsetup{justification=justified}
    \caption{(a) Trapping events occurred for each excitation intensity, classified as taking place at Hotspot 1 (blue line), Hotspot 2 (red line), or undetermined (purple line). The numbers in blue, red, and purple correspond to the number of trapping events, respectively. (b) Simulated temperature rise at the center of illumination on the gold film as a function of the incident laser intensity. There is a temperature rise of 10.6$^o$C for every 1~mW/$\mu\mbox{m}^2$ intensity increase.}
    \label{Events-Thermal}
\end{figure}

From Figure~\ref{Hist_k_all}, we see that many more particles are trapped with low stiffness values, and that, may seem like corresponding to trapping events at Hotspot 2. This finding is unexpected considering the availability of hotspots for trapping, there being twice as many of type 1 than type 2. To investigate this deviation, we plot the number of trapping events for each excitation intensity in Figure~\ref{Events_I}. The numbers in blue and red show the events for Hotspot 1 and 2, respectively, for two intensity regions. We see that, until an intensity of about 1.5~mW/$\mu\mbox{m}^2$, there was a ratio of about 2/1 for Hotspot 1/Hotspot 2 trapping events, as expected. For higher intensities, we observed a very large number of trapping events with very low trap stiffnesses. These trapping events, shown in purple in Figure~\ref{Events_I}, cannot be classified as occurring in Hotspot 1 or 2 based on their stiffness values alone. Our hypothesis is that the origin of their low measured stiffness values is not due to a weak optical force but rather the effect of an additional destabilization force. Nevertheless, even beyond 1.5~mW/$\mu\mbox{m}^2$, there were some trapping events that clearly occurred in Hotspot~1 (see Figure~\ref{Events_I}), due to their characteristically high stiffness values.

To determine other effects that could reduce the trap stiffness, we considered plasmonic heating and the subsequent temperature rise of the surrounding environment (water) as we increased the laser intensity. In Figure~\ref{Thermal-I}, we simulated the temperature rise at the center of illumination on the metamaterial structure, with parameters matching the experimental conditions. A linear relationship is observed as expected from theory~\cite{Roxworthy2014Convection}, showing a rise of about 10.6$^o$C per 1~mW/$\mu\mbox{m}^2$ intensity increase. It is worth noting that the convection flows and thermophoretic forces are proportional to the temperature rise, $\Delta T$, and its gradient, $\nabla T$. The temperature gradient, $\nabla T$, is related to factors such as the thermal conductivity of the trapped particle, water solution, and metal nanostructure surfaces. Due to the significant mismatch in thermal conductivity between the gold boundaries’ surfaces and the water solution within the hotspot regions, large temperature gradients, $\nabla T$, of approximately 40$^o$C/$\mu$m can be observed per 1~mW/$\mu\mbox{m}^2$ intensity increase. At 1.5~mW/$\mu\mbox{m}^2$, there is a temperature rise, $\Delta T$, of about 15.9$^o$C and a temperature gradient, $\nabla T$, of about 60$^o\mbox{C}/\mu\mbox{m}$, which are sufficient to generate strong convection currents and thermophoretic forces that hinder efficient trapping~\cite{Roxworthy2014Convection, Kotsifaki2022Thermal}. The temperature for which thermal effects start contributing (positively or negatively) to the trapping process depends on many parameters such as the gold film deposition (thickness, adhesion layer, substrate)~\cite{Roxworthy2014Convection}, the height of the microwell where the particles are suspended~\cite{Braibanti, Xu2019AlldielectricNF}, the Soret coefficient of the nanoparticles and their interaction with surfactants~\cite{Jiang2020QuantifyingTR, Kotsifaki2022Thermal}, the plasmonic design~\cite{Baffou2013ThermoArrays, Cichos2022}, and so on. Hence, controlling the trapping landscape as the incident intensity increases is very complicated and, therefore, it is highly desirable to trap with low intensities. 

Considering our experimental conditions, an incident laser intensity lower than 1.5~mW/$\mu\mbox{m}^2$ should be used. The temperature change at this intensity (15.9$^o$C) is typical in similar experiments~\cite{Wenger2019TemperatureMeasur, Kotsifaki2022Thermal}. Above this temperature, the temperature gradients created strong convection flows in the solution that facilitated fast transport of thermophilic particles toward the hottest areas~\cite{WurgerSoret, Kotnala2019Overcomingdiffusion, Bouloumis2020FastAE, Jiang2020QuantifyingTR}, leading to fast trapping of a large number of particles (see Figure~\ref{Events_I}); the convection flows transported particles close to the hotspots from a distance further than the optical forces range. However, due to their momentum and kinetic energy on approaching the hotspots, the particles were harder to trap, leading to lower stiffness values. This affected particles trapped in both hotspots and, therefore, we cannot determine the trapping location (Hotspot 1 or 2) by the measured stiffness value alone for these (purple) trapping events.

\begin{figure}[ht!]
     \centering 
     \captionsetup{justification=raggedright,singlelinecheck=false}
     \begin{subfigure}[b]{0.48\textwidth}
         \centering
         \caption{}
         \includegraphics[width=\textwidth]{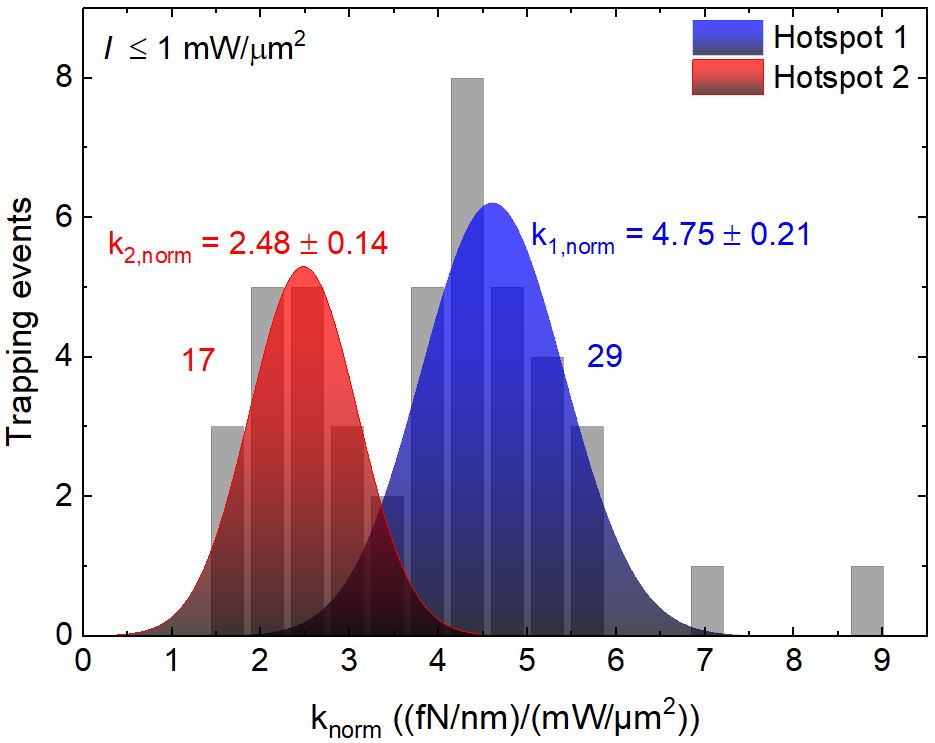}
         \label{Hist_k_1}
     \end{subfigure}
    \hfill
     \begin{subfigure}[b]{0.49\textwidth}
         \centering
         \caption{}
         \includegraphics[ width=\textwidth]{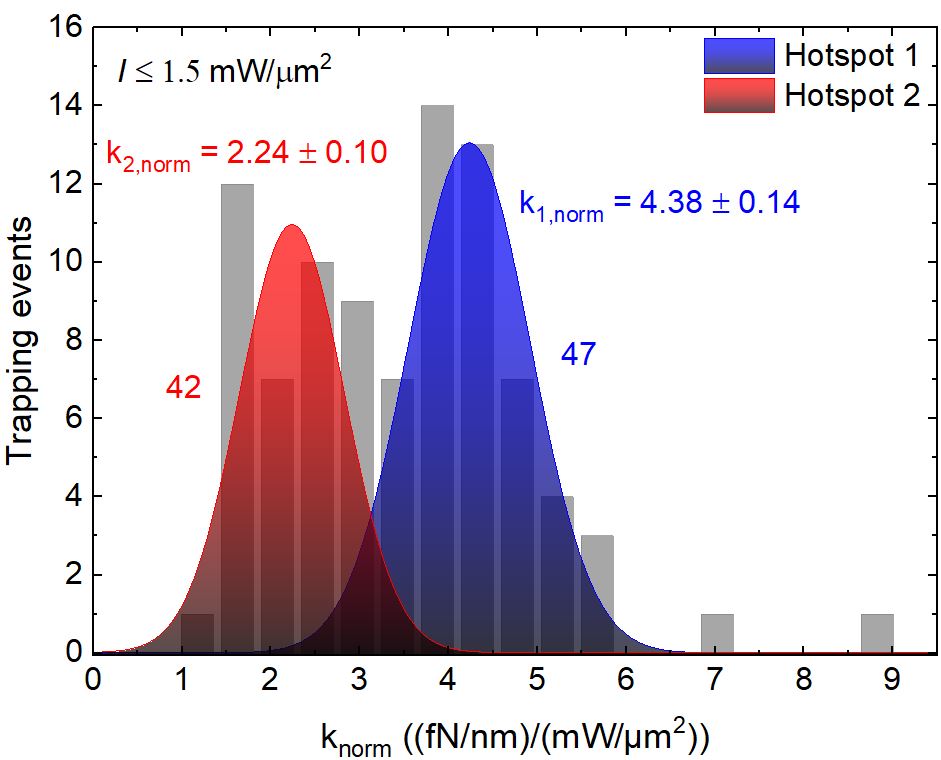}
         \label{Hist_k_1.5}
     \end{subfigure}
    \captionsetup{justification=justified}
    \caption{Histograms of the measured normalized trap stiffnesses when we include trapping events with excitation intensities (\textit{I}) of up to (a) 1~mW/$\mu\mbox{m}^2$ and (b) 1.5~mW/$\mu\mbox{m}^2$ with Gaussian fits on the hotspot distributions. The values $k_{1,norm}$ and $k_{2,norm}$ refer to the average stiffness value for particles trapped in Hotspot 1 (blue curve) and Hotspot 2 (red curve), respectively. The numbers in blue and red refer to the number of trapping events.}
    \label{Hist_k_Low_Intensities}
\end{figure}

The effect of temperature rise may also come into play at even lower intensities. In Figure~\ref{Hist_k_Low_Intensities}, we plot the trap stiffness histograms for trapping events with laser intensities up to 1~mW/$\mu\mbox{m}^2$ (Figure~\ref{Hist_k_1}) and up to 1.5~mW/$\mu\mbox{m}^2$ (Figure~\ref{Hist_k_1.5}). Interestingly, we see that for intensities lower than 1~mW/$\mu\mbox{m}^2$, the distributions of events over the two hotspots are distinguishable, with more trapping events at Hotspot 1, as expected. However, even before the incident intensity exceeds 1.5~mW/$\mu\mbox{m}^2$, we see that it is harder to distinguish between the two distributions (Figure~\ref{Hist_k_1.5}) implying that heating effects were already present. In addition, looking at the average stiffness values ($k_{1,norm}$ and $k_{2,norm}$), we note that, as we include higher intensities, the average value decreases for both hotspots owing to thermal effects. Therefore, using an intensity of less than 1~mW/$\mu\mbox{m}^2$ should be a safe regime to trap nanoparticles efficiently.

\begin{figure}[ht!]
     \centering 
         \includegraphics[width=0.6\textwidth]{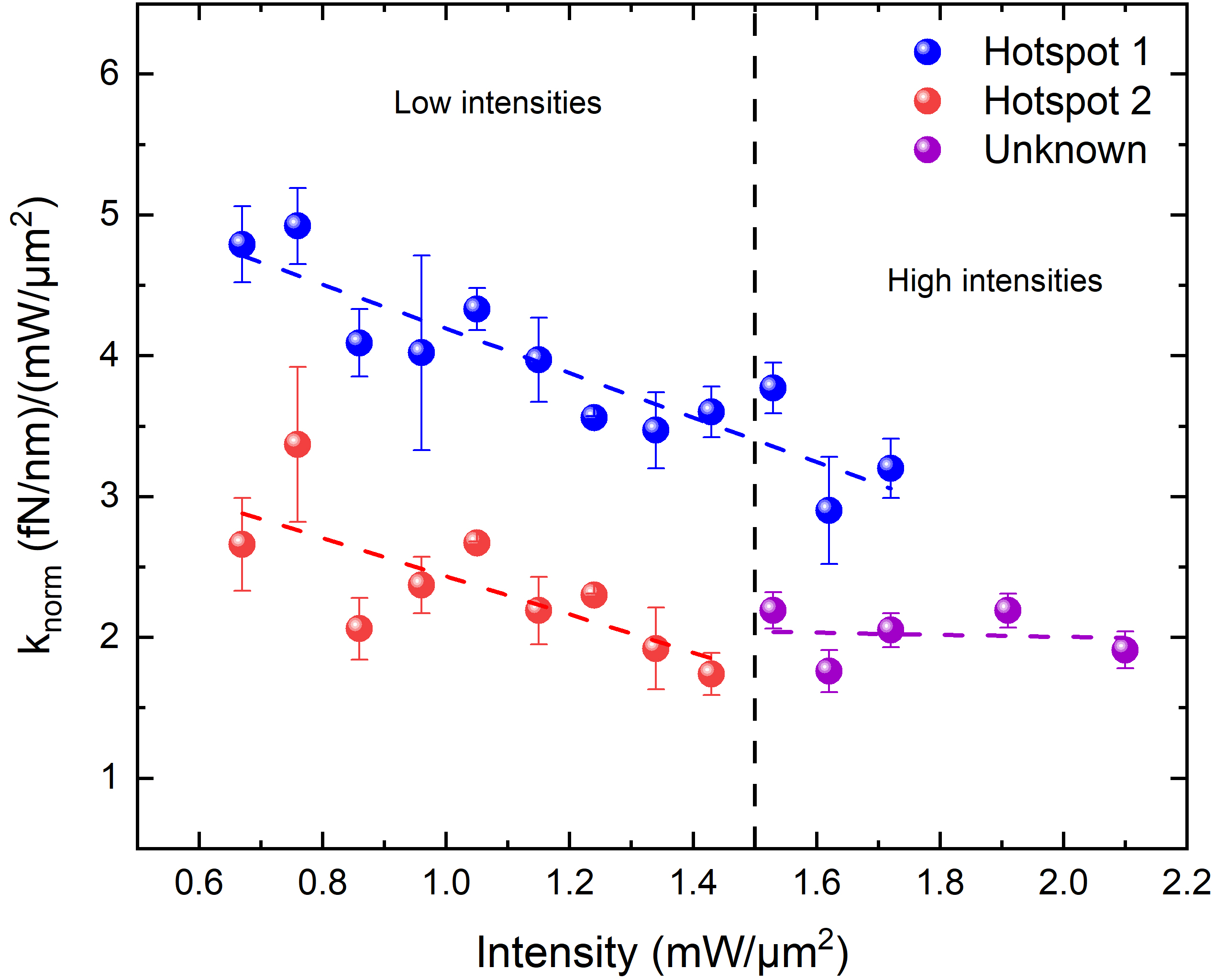}
        \caption[Normalized trap stiffness as a function of the excitation laser intensity]{Normalized trap stiffnesses as a function of the excitation laser intensity for the two hotspots (blue and red), and for the trapping events which cannot be classified in one or other of the hotspots due to the low values (purple).}
        \label{k-I}
\end{figure}

Based on the theory of optical forces, the trap stiffness increases linearly with the intensity~\cite{jones_marago_volpe_2015}. Thus, the normalized trap stiffness (trap stiffness divided by the intensity) should be constant as a function of intensity if only optical forces are present. In our case, as shown in Figure~\ref{k-I}, the normalized trap stiffness linearly decreases with increasing intensity by about 1.5~(fN/nm)/(mW/$\mu\mbox{m}^2$) per 1~mW/$\mu\mbox{m}^2$ for both hotspots. Evidently, the temperature increase and the resulting thermal flows counteract the strong optical forces, with both phenomena linearly correlated to the incident intensity, therefore resulting in  inefficient trapping. It worth noting once again that given the experimental conditions in this work, working with as low an intensity as possible leads to the most efficient trapping. That may also be due to self-induced back-action (SIBA)-assisted trapping since we excited the plasmonic field with a near-resonant wavelength~\cite{Mestres2016Unravelling, BouloumisEnabling}. Finally, an interesting observation is that for those trapping events at high laser intensities, leading to low stiffness values (see purple points in Figure~\ref{k-I}), the normalized trap stiffness remained constant with increasing intensity. This could be due to thermal effects and optical forces reaching an equilibrium beyond a critical temperature. Investigating the trapping process in this high intensities regime is an interesting topic for future study. 

\section{4. Conclusions}

We demonstrated trapping of custom-synthesized nanoparticles based on molecular GQDs having quasi-core/shell structure, with fluorescent D7H nanographene molecules inside the core and hydrophilic moieties of amphiphilic polymers as the shell, and a total diameter of around 20~nm, using metamaterial plasmonic tweezers. Such molecular GQDs have previously been studied for lysosome-targeted cancer photodynamic therapy, exhibiting their potential in medical applications. In this work, our goal was to study their trapping conditions and positioning dynamics on a periodic metamaterial array for potential application as single-photon sources in the field of quantum technologies. Multiple trapping events were achieved with a range of incident laser intensities. The stiffness values measured were very high ranging between 0.47 - 8.8~(fN/nm)/(mW/$\mu\mbox{m}^2$) and, based on statistical analysis, the hotspot location (Hotspot 1 or 2) where a particle was trapped could be determined. However, we noticed a discrepancy in the expected particle trapping distribution, which was caused by the temperature rise and induced thermal flows in the solution as the incident laser intensity increased. The method  used to classify the trapping hotspot locations only considered optical forces and, thus, is insufficient beyond an incident intensity of about 1.5~mW/$\mu\mbox{m}^2$ where thermal effects come into play.  At this intensity, the temperature rise was about 15.9$^o$C, based on simulations. Beyond this temperature we observed fast transport of particles towards the hotspots, and the resulting trap stiffness values were low due to destabilization of the trapping mechanism. Further analysis showed that, even for intensities lower than 1.5~mW/$\mu\mbox{m}^2$, thermal effects appeared, and trapping with intensities lower than about 1~mW/$\mu\mbox{m}^2$ is desirable if one wants to completely avoid thermal effects. Investigating the trapping performance beyond the critical temperature is an interesting topic since our data imply that beyond this critical value an equilibrium is reached between the optical and thermal forces. Being able to control the trapping process in this high intensity/temperature regime may be useful for nanopositioning of quantum emitters in an ordered  array  for potential use as scalable single-photon sources  and studying collective radiative interactions.

\newpage
\section{TOC Graphic}

\begin{figure}
    \centering
    \includegraphics{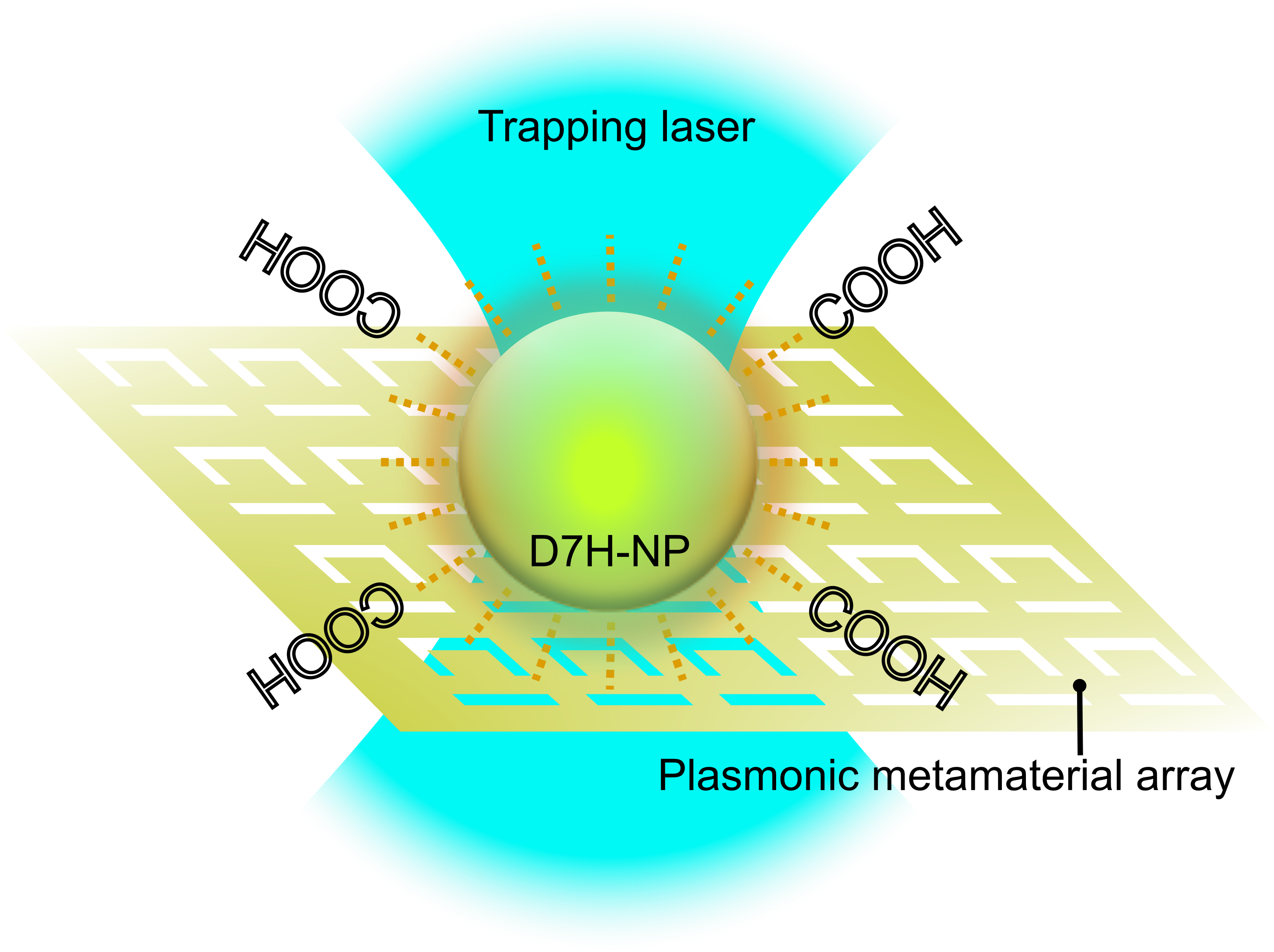}
    \captionsetup{labelformat=empty}
    \caption{Trapping of a D7H-NP on a plasmonic metamaterial array using near-field forces.}
\end{figure}

\newpage

\subsection{Authors Information}

\textbf{Theodoros D. Bouloumis} $^{}$\orcidA{}:0000-0002-5264-7338,\\
Light – Matter Interactions for Quantum Technologies Unit, Okinawa Institute of Science and Technology Graduate University, 1919-1 Tancha, Onna-son, Okinawa, 904-0495, Japan, theodoros.bouloumis@oist.jp\\

\noindent \textbf{Hao Zhao} $^{}$\orcidB{}:0000-0002-9125-617X,\\
Organic and Carbon Nanomaterials Unit, Okinawa Institute of Science and Technology Graduate University, 1919-1 Tancha, Onna-son, Okinawa, 904-0495, Japan, hao.zhao@oist.jp\\

\noindent \textbf{Nikolaos Kokkinidis} $^{}$\orcidC{}:0009-0007-2479-975X,\\
Light – Matter Interactions for Quantum Technologies Unit, Okinawa Institute of Science and Technology Graduate University, 1919-1 Tancha, Onna-son, Okinawa, 904-0495, Japan, nikolaos.kokkinidis@oist.jp\\

\noindent \textbf{Yunbin Hu} $^{}$\orcidD{}:0000-0001-5346-7059,\\
College of Chemistry and Chemical Engineering, Central South University, Changsha, 410083, China, huyunbin@csu.edu.cn\\

\noindent \textbf{Viet Giang Truong} $^{}$\orcidE{}:0000-0003-3589-7850,\\
Light – Matter Interactions for Quantum Technologies Unit, Okinawa Institute of Science and Technology Graduate University, 1919-1 Tancha, Onna-son, Okinawa, 904-0495, Japan, v.g.truong@oist.jp\\

\noindent \textbf{Akimitsu Narita} $^{}$\orcidF{}:0000-0002-3625-522X,\\
Organic and Carbon Nanomaterials Unit, Okinawa Institute of Science and Technology Graduate University, 1919-1 Tancha, Onna-son, Okinawa, 904-0495, Japan, akimitsu.narita@oist.jp\\

\noindent \textbf{S\'ile {Nic Chormaic}} $^{}$\orcidG{}:0000-0003-4276-2014,\\
Light – Matter Interactions for Quantum Technologies Unit, Okinawa Institute of Science and Technology Graduate University, 1919-1 Tancha, Onna-son, Okinawa, 904-0495, Japan, sile.nicchormaic@oist.jp

\noindent The authors declare no competing financial interests.

\begin{acknowledgement}

This work was supported by the Okinawa Institute of Science and Technology Graduate University (OIST), JSPS KAKENHI Grant No. 22KJ3083 (T.D. Bouloumis) and 23KF0075 (A. Narita), and Natural Science Foundation of China (No.22375219; Y. Hu). T.D. Bouloumis acknowledges the JSPS DC2/PD Research Fellowship and H. Zhao acknowledges the JSPS Postdoctoral Fellowship for Research in Japan. V.G. Truong acknowledges support by JSPS KAKENHI (Grant-in-Aid for Scientific Research), Grant No. JP23K04618. The authors thank M. Ozer, M. Z. Jalaludeen, L. Zhou, the Scientific Computing and Data Analysis Section, and the Engineering Section of OIST for support.

\end{acknowledgement}

\newpage

\newpage
\bibliography{D7H-Manuscript}

\providecommand{\latin}[1]{#1}
\makeatletter
\providecommand{\doi}
  {\begingroup\let\do\@makeother\dospecials
  \catcode`\{=1 \catcode`\}=2 \doi@aux}
\providecommand{\doi@aux}[1]{\endgroup\texttt{#1}}
\makeatother
\providecommand*\mcitethebibliography{\thebibliography}
\csname @ifundefined\endcsname{endmcitethebibliography}
  {\let\endmcitethebibliography\endthebibliography}{}
\begin{mcitethebibliography}{64}
\providecommand*\natexlab[1]{#1}
\providecommand*\mciteSetBstSublistMode[1]{}
\providecommand*\mciteSetBstMaxWidthForm[2]{}
\providecommand*\mciteBstWouldAddEndPuncttrue
  {\def\EndOfBibitem{\unskip.}}
\providecommand*\mciteBstWouldAddEndPunctfalse
  {\let\EndOfBibitem\relax}
\providecommand*\mciteSetBstMidEndSepPunct[3]{}
\providecommand*\mciteSetBstSublistLabelBeginEnd[3]{}
\providecommand*\EndOfBibitem{}
\mciteSetBstSublistMode{f}
\mciteSetBstMaxWidthForm{subitem}{(\alph{mcitesubitemcount})}
\mciteSetBstSublistLabelBeginEnd
  {\mcitemaxwidthsubitemform\space}
  {\relax}
  {\relax}

\bibitem[Novotny \latin{et~al.}(1997)Novotny, Bian, and
  Xie]{Novotny1997Nanometric}
Novotny,~L.; Bian,~R.~X.; Xie,~X.~S. Theory of nanometric optical tweezers.
  \emph{Physical Review Letters} \textbf{1997}, \emph{79}, 645--648\relax
\mciteBstWouldAddEndPuncttrue
\mciteSetBstMidEndSepPunct{\mcitedefaultmidpunct}
{\mcitedefaultendpunct}{\mcitedefaultseppunct}\relax
\EndOfBibitem
\bibitem[Murray \latin{et~al.}(2000)Murray, Kagan, and Bawendi]{Kagan2000QDs}
Murray,~C.~B.; Kagan,~C.~R.; Bawendi,~M.~G. Synthesis and characterization of
  monodisperse nanocrystals and close-packed nanocrystal assemblies.
  \emph{Annual Review of Materials Science} \textbf{2000}, \emph{30},
  545--610\relax
\mciteBstWouldAddEndPuncttrue
\mciteSetBstMidEndSepPunct{\mcitedefaultmidpunct}
{\mcitedefaultendpunct}{\mcitedefaultseppunct}\relax
\EndOfBibitem
\bibitem[Cheng \latin{et~al.}(2014)Cheng, Lowe, Reece, and
  Gooding]{Reece2014ColloidalQD}
Cheng,~X.; Lowe,~S.~B.; Reece,~P.~J.; Gooding,~J.~J. Colloidal silicon quantum
  dots: from preparation to the modification of self-assembled monolayers
  (SAMs) for bio-applications. \emph{Chemical Society Reviews} \textbf{2014},
  \emph{43}, 2680--2700\relax
\mciteBstWouldAddEndPuncttrue
\mciteSetBstMidEndSepPunct{\mcitedefaultmidpunct}
{\mcitedefaultendpunct}{\mcitedefaultseppunct}\relax
\EndOfBibitem
\bibitem[Kagan \latin{et~al.}(2021)Kagan, Bassett, Murray, and
  Thompson]{Thompson2021QuantumInformation}
Kagan,~C.~R.; Bassett,~L.~C.; Murray,~C.~B.; Thompson,~S.~M. Colloidal quantum
  dots as platforms for quantum information science. \emph{Chemical Reviews}
  \textbf{2021}, \emph{121}, 3186--3233, doi: 10.1021/acs.chemrev.0c00831\relax
\mciteBstWouldAddEndPuncttrue
\mciteSetBstMidEndSepPunct{\mcitedefaultmidpunct}
{\mcitedefaultendpunct}{\mcitedefaultseppunct}\relax
\EndOfBibitem
\bibitem[Aharonovich \latin{et~al.}(2011)Aharonovich, Greentree, and
  Prawer]{AharonovichDiamond2011}
Aharonovich,~I.; Greentree,~A.~D.; Prawer,~S. Diamond photonics. \emph{Nature
  Photonics} \textbf{2011}, \emph{5}, 397--405\relax
\mciteBstWouldAddEndPuncttrue
\mciteSetBstMidEndSepPunct{\mcitedefaultmidpunct}
{\mcitedefaultendpunct}{\mcitedefaultseppunct}\relax
\EndOfBibitem
\bibitem[Reineck \latin{et~al.}(2019)Reineck, Trindade, Havlik, Stursa,
  Heffernan, Elbourne, Orth, Capelli, Cigler, Simpson, and
  Gibson]{Gibson2019Nanodiamonds}
Reineck,~P.; Trindade,~L.~F.; Havlik,~J.; Stursa,~J.; Heffernan,~A.;
  Elbourne,~A.; Orth,~A.; Capelli,~M.; Cigler,~P.; Simpson,~D.~A.;
  Gibson,~B.~C. Not all fluorescent nanodiamonds are created equal: A
  comparative study. \emph{Particle \& Particle Systems Characterization}
  \textbf{2019}, \emph{36}, 1900009,
  https://doi.org/10.1002/ppsc.201900009\relax
\mciteBstWouldAddEndPuncttrue
\mciteSetBstMidEndSepPunct{\mcitedefaultmidpunct}
{\mcitedefaultendpunct}{\mcitedefaultseppunct}\relax
\EndOfBibitem
\bibitem[Jin \latin{et~al.}(2015)Jin, Owour, Lei, and Ge]{JIN2015439}
Jin,~Z.; Owour,~P.; Lei,~S.; Ge,~L. Graphene, graphene quantum dots and their
  applications in optoelectronics. \emph{Current Opinion in Colloid \&
  Interface Science} \textbf{2015}, \emph{20}, 439--453\relax
\mciteBstWouldAddEndPuncttrue
\mciteSetBstMidEndSepPunct{\mcitedefaultmidpunct}
{\mcitedefaultendpunct}{\mcitedefaultseppunct}\relax
\EndOfBibitem
\bibitem[Yan \latin{et~al.}(2019)Yan, Gong, Chen, Zeng, Huang, Pu, Liu, and
  Chen]{YiboGQD2019}
Yan,~Y.; Gong,~J.; Chen,~J.; Zeng,~Z.; Huang,~W.; Pu,~K.; Liu,~J.; Chen,~P.
  Recent Advances on Graphene Quantum Dots: From Chemistry and Physics to
  Applications. \emph{Advanced Materials} \textbf{2019}, \emph{31},
  1808283\relax
\mciteBstWouldAddEndPuncttrue
\mciteSetBstMidEndSepPunct{\mcitedefaultmidpunct}
{\mcitedefaultendpunct}{\mcitedefaultseppunct}\relax
\EndOfBibitem
\bibitem[Toninelli \latin{et~al.}(2021)Toninelli, Gerhardt, Clark,
  Reserbat-Plantey, Götzinger, Ristanović, Colautti, Lombardi, Major,
  Deperasińska, Pernice, Koppens, Kozankiewicz, Gourdon, Sandoghdar, and
  Orrit]{Toninenlli2021OrganMol}
Toninelli,~C. \latin{et~al.}  Single organic molecules for photonic quantum
  technologies. \emph{Nature Materials} \textbf{2021}, \emph{20},
  1615--1628\relax
\mciteBstWouldAddEndPuncttrue
\mciteSetBstMidEndSepPunct{\mcitedefaultmidpunct}
{\mcitedefaultendpunct}{\mcitedefaultseppunct}\relax
\EndOfBibitem
\bibitem[Gaither-Ganim \latin{et~al.}(2023)Gaither-Ganim, Newlon, Anderson, and
  Lee]{Gaither2023OrgMolec}
Gaither-Ganim,~M.~B.; Newlon,~S.~A.; Anderson,~M.~G.; Lee,~B. Organic molecule
  single-photon sources. \emph{Oxford Open Materials Science} \textbf{2023},
  \emph{3}, itac017\relax
\mciteBstWouldAddEndPuncttrue
\mciteSetBstMidEndSepPunct{\mcitedefaultmidpunct}
{\mcitedefaultendpunct}{\mcitedefaultseppunct}\relax
\EndOfBibitem
\bibitem[Protesescu \latin{et~al.}(2015)Protesescu, Yakunin, Bodnarchuk, Krieg,
  Caputo, Hendon, Yang, Walsh, and Kovalenko]{Maksym2015PerovskiteQDs}
Protesescu,~L.; Yakunin,~S.; Bodnarchuk,~M.~I.; Krieg,~F.; Caputo,~R.;
  Hendon,~C.~H.; Yang,~R.~X.; Walsh,~A.; Kovalenko,~M.~V. Nanocrystals of
  cesium lead halide perovskites (CsPbX3, X = Cl, Br, and I): Novel
  optoelectronic materials showing bright emission with wide color gamut.
  \emph{Nano Letters} \textbf{2015}, \emph{15}, 3692--3696, doi:
  10.1021/nl5048779\relax
\mciteBstWouldAddEndPuncttrue
\mciteSetBstMidEndSepPunct{\mcitedefaultmidpunct}
{\mcitedefaultendpunct}{\mcitedefaultseppunct}\relax
\EndOfBibitem
\bibitem[Freitas \latin{et~al.}(2020)Freitas, Neves, Nouws, and
  Delerue-Matos]{Matos2020QDNanolabels}
Freitas,~M.; Neves,~M. M. P.~S.; Nouws,~H. P.~A.; Delerue-Matos,~C. Quantum
  dots as nanolabels for breast cancer biomarker HER2-ECD analysis in human
  serum. \emph{Talanta} \textbf{2020}, \emph{208}, 120430\relax
\mciteBstWouldAddEndPuncttrue
\mciteSetBstMidEndSepPunct{\mcitedefaultmidpunct}
{\mcitedefaultendpunct}{\mcitedefaultseppunct}\relax
\EndOfBibitem
\bibitem[Zhao \latin{et~al.}(2022)Zhao, Xu, Zhou, Hu, Huang, and
  Narita]{Hao2022D7H}
Zhao,~H.; Xu,~X.; Zhou,~L.; Hu,~Y.; Huang,~Y.; Narita,~A. Water-soluble
  nanoparticles with twisted double [7]carbohelicene for lysosome-targeted
  cancer photodynamic therapy. \emph{Small} \textbf{2022}, \emph{18},
  2105365\relax
\mciteBstWouldAddEndPuncttrue
\mciteSetBstMidEndSepPunct{\mcitedefaultmidpunct}
{\mcitedefaultendpunct}{\mcitedefaultseppunct}\relax
\EndOfBibitem
\bibitem[Jang \latin{et~al.}(2010)Jang, Jun, Jang, Lim, Kim, and
  Kim]{Jang2010QDWhiteLED}
Jang,~E.; Jun,~S.; Jang,~H.; Lim,~J.; Kim,~B.; Kim,~Y. White-light-emitting
  diodes with quantum dot color converters for display backlights.
  \emph{Advanced Materials} \textbf{2010}, \emph{22}, 3076--3080,
  https://doi.org/10.1002/adma.201000525\relax
\mciteBstWouldAddEndPuncttrue
\mciteSetBstMidEndSepPunct{\mcitedefaultmidpunct}
{\mcitedefaultendpunct}{\mcitedefaultseppunct}\relax
\EndOfBibitem
\bibitem[Liu \latin{et~al.}(2020)Liu, Lin, Hyun, Sher, Lv, Luo, Jiang, Wu, Ho,
  Kuo, and He]{Liu2020QDLED}
Liu,~Z.; Lin,~C.-H.; Hyun,~B.-R.; Sher,~C.-W.; Lv,~Z.; Luo,~B.; Jiang,~F.;
  Wu,~T.; Ho,~C.-H.; Kuo,~H.-C.; He,~J.-H. Micro-light-emitting diodes with
  quantum dots in display technology. \emph{Light: Science \& Applications}
  \textbf{2020}, \emph{9}, 83\relax
\mciteBstWouldAddEndPuncttrue
\mciteSetBstMidEndSepPunct{\mcitedefaultmidpunct}
{\mcitedefaultendpunct}{\mcitedefaultseppunct}\relax
\EndOfBibitem
\bibitem[Kagan \latin{et~al.}(2016)Kagan, Lifshitz, Sargent, and
  Talapin]{Kagan2016QDDevices}
Kagan,~C.~R.; Lifshitz,~E.; Sargent,~E.~H.; Talapin,~D.~V. Building devices
  from colloidal quantum dots. \emph{Science} \textbf{2016}, \emph{353},
  aac5523, doi: 10.1126/science.aac5523\relax
\mciteBstWouldAddEndPuncttrue
\mciteSetBstMidEndSepPunct{\mcitedefaultmidpunct}
{\mcitedefaultendpunct}{\mcitedefaultseppunct}\relax
\EndOfBibitem
\bibitem[Takekuma \latin{et~al.}(2021)Takekuma, Leng, Tateishi, Xu, Chan,
  Ryuzaki, Wang, Okamoto, and Tamada]{TakeQD2021}
Takekuma,~H.; Leng,~J.; Tateishi,~K.; Xu,~Y.; Chan,~Y.; Ryuzaki,~S.; Wang,~P.;
  Okamoto,~K.; Tamada,~K. Layer number-dependent enhanced photoluminescence
  from a quantum dot metamaterial optical resonator. \emph{ACS Applied
  Electronic Materials} \textbf{2021}, \emph{3}, 468--475\relax
\mciteBstWouldAddEndPuncttrue
\mciteSetBstMidEndSepPunct{\mcitedefaultmidpunct}
{\mcitedefaultendpunct}{\mcitedefaultseppunct}\relax
\EndOfBibitem
\bibitem[Ha \latin{et~al.}(2021)Ha, Hur, Pathak, Jeong, and Woo]{Ha2021Organic}
Ha,~J.~M.; Hur,~S.~H.; Pathak,~A.; Jeong,~J.-E.; Woo,~H.~Y. Recent advances in
  organic luminescent materials with narrowband emission. \emph{NPG Asia
  Materials} \textbf{2021}, \emph{13}, 53\relax
\mciteBstWouldAddEndPuncttrue
\mciteSetBstMidEndSepPunct{\mcitedefaultmidpunct}
{\mcitedefaultendpunct}{\mcitedefaultseppunct}\relax
\EndOfBibitem
\bibitem[Lounis and Moerner(2000)Lounis, and Moerner]{Lounis2000Molecule}
Lounis,~B.; Moerner,~W.~E. Single photons on demand from a single molecule at
  room temperature. \emph{Nature} \textbf{2000}, \emph{407}, 491--493\relax
\mciteBstWouldAddEndPuncttrue
\mciteSetBstMidEndSepPunct{\mcitedefaultmidpunct}
{\mcitedefaultendpunct}{\mcitedefaultseppunct}\relax
\EndOfBibitem
\bibitem[Zhang \latin{et~al.}(2008)Zhang, Dang, Urabe, Wang, Sun, and
  Nurmikko]{Zhang:08}
Zhang,~Q.; Dang,~C.; Urabe,~H.; Wang,~J.; Sun,~S.; Nurmikko,~A. Large ordered
  arrays of single photon sources based on II--VI semiconductor colloidal
  quantum dot. \emph{Opt. Express} \textbf{2008}, \emph{16}, 19592--19599\relax
\mciteBstWouldAddEndPuncttrue
\mciteSetBstMidEndSepPunct{\mcitedefaultmidpunct}
{\mcitedefaultendpunct}{\mcitedefaultseppunct}\relax
\EndOfBibitem
\bibitem[Mizuochi \latin{et~al.}(2012)Mizuochi, Makino, Kato, Takeuchi, Ogura,
  Okushi, Nothaft, Neumann, Gali, Jelezko, Wrachtrup, and
  Yamasaki]{Mizu2012diamond}
Mizuochi,~N.; Makino,~T.; Kato,~H.; Takeuchi,~D.; Ogura,~M.; Okushi,~H.;
  Nothaft,~M.; Neumann,~P.; Gali,~A.; Jelezko,~F.; Wrachtrup,~J.; Yamasaki,~S.
  Electrically driven single-photon source at room temperature in diamond.
  \emph{Nature Photonics} \textbf{2012}, \emph{6}, 299--303\relax
\mciteBstWouldAddEndPuncttrue
\mciteSetBstMidEndSepPunct{\mcitedefaultmidpunct}
{\mcitedefaultendpunct}{\mcitedefaultseppunct}\relax
\EndOfBibitem
\bibitem[Gao \latin{et~al.}(2015)Gao, Imamoglu, Bernien, and
  Hanson]{Imamoglu2015ReviewEntanglement}
Gao,~W.~B.; Imamoglu,~A.; Bernien,~H.; Hanson,~R. Coherent manipulation,
  measurement and entanglement of individual solid-state spins using optical
  fields. \emph{Nature Photonics} \textbf{2015}, \emph{9}, 363--373\relax
\mciteBstWouldAddEndPuncttrue
\mciteSetBstMidEndSepPunct{\mcitedefaultmidpunct}
{\mcitedefaultendpunct}{\mcitedefaultseppunct}\relax
\EndOfBibitem
\bibitem[Eich \latin{et~al.}(2022)Eich, Spiekermann, Gehring, Sommer, Bankwitz,
  Schrinner, Preuß, Michaelis~de Vasconcellos, Bratschitsch, Pernice, and
  Schuck]{Carsten2022Singlephoton}
Eich,~A.; Spiekermann,~T.~C.; Gehring,~H.; Sommer,~L.; Bankwitz,~J.~R.;
  Schrinner,~P. P.~J.; Preuß,~J.~A.; Michaelis~de Vasconcellos,~S.;
  Bratschitsch,~R.; Pernice,~W. H.~P.; Schuck,~C. Single-photon emission from
  individual nanophotonic-integrated colloidal quantum dots. \emph{ACS
  Photonics} \textbf{2022}, \emph{9}, 551--558, doi:
  10.1021/acsphotonics.1c01493\relax
\mciteBstWouldAddEndPuncttrue
\mciteSetBstMidEndSepPunct{\mcitedefaultmidpunct}
{\mcitedefaultendpunct}{\mcitedefaultseppunct}\relax
\EndOfBibitem
\bibitem[Horowitz \latin{et~al.}(2012)Horowitz, Alemán, Christle, Cleland, and
  Awschalom]{Horowitz2012ElectronSR}
Horowitz,~V.~R.; Alemán,~B.~J.; Christle,~D.~J.; Cleland,~A.~N.;
  Awschalom,~D.~D. Electron spin resonance of nitrogen-vacancy centers in
  optically trapped nanodiamonds. \emph{Proceedings of the National Academy of
  Sciences} \textbf{2012}, \emph{109}, 13493--13497, doi:
  10.1073/pnas.1211311109\relax
\mciteBstWouldAddEndPuncttrue
\mciteSetBstMidEndSepPunct{\mcitedefaultmidpunct}
{\mcitedefaultendpunct}{\mcitedefaultseppunct}\relax
\EndOfBibitem
\bibitem[Jauffred \latin{et~al.}(2014)Jauffred, Kyrsting, Arnspang, Reihani,
  and Oddershede]{Jauffred2014Subdiffraction}
Jauffred,~L.; Kyrsting,~A.; Arnspang,~E.~C.; Reihani,~S. N.~S.;
  Oddershede,~L.~B. Sub-diffraction positioning of a two-photon excited and
  optically trapped quantum dot. \emph{Nanoscale} \textbf{2014}, \emph{6},
  6997--7003\relax
\mciteBstWouldAddEndPuncttrue
\mciteSetBstMidEndSepPunct{\mcitedefaultmidpunct}
{\mcitedefaultendpunct}{\mcitedefaultseppunct}\relax
\EndOfBibitem
\bibitem[Jauffred \latin{et~al.}(2008)Jauffred, Richardson, and
  Oddershede]{Liselotte2008QD}
Jauffred,~L.; Richardson,~A.~C.; Oddershede,~L.~B. Three-dimensional optical
  control of individual quantum dots. \emph{Nano Letters} \textbf{2008},
  \emph{8}, 3376--80\relax
\mciteBstWouldAddEndPuncttrue
\mciteSetBstMidEndSepPunct{\mcitedefaultmidpunct}
{\mcitedefaultendpunct}{\mcitedefaultseppunct}\relax
\EndOfBibitem
\bibitem[Li \latin{et~al.}(2022)Li, Li, Qiu, and Li]{Li2022PhotothermalQD}
Li,~J.; Li,~Z.; Qiu,~J.; Li,~J. Photothermal optimization of quantum dot
  converters for high-power solid-state light sources. \emph{Advanced Optical
  Materials} \textbf{2022}, \emph{10}, 2102201\relax
\mciteBstWouldAddEndPuncttrue
\mciteSetBstMidEndSepPunct{\mcitedefaultmidpunct}
{\mcitedefaultendpunct}{\mcitedefaultseppunct}\relax
\EndOfBibitem
\bibitem[Jauffred and Oddershede(2010)Jauffred, and
  Oddershede]{Liselotte2010TwophotonQD}
Jauffred,~L.; Oddershede,~L.~B. Two-photon quantum dot excitation during
  optical trapping. \emph{Nano Letters} \textbf{2010}, \emph{10},
  1927--30\relax
\mciteBstWouldAddEndPuncttrue
\mciteSetBstMidEndSepPunct{\mcitedefaultmidpunct}
{\mcitedefaultendpunct}{\mcitedefaultseppunct}\relax
\EndOfBibitem
\bibitem[Tsuboi \latin{et~al.}(2010)Tsuboi, Shoji, Kitamura, Takase, Murakoshi,
  Mizumoto, and Ishihara]{Tsuboi2010QDGap}
Tsuboi,~Y.; Shoji,~T.; Kitamura,~N.; Takase,~M.; Murakoshi,~K.; Mizumoto,~Y.;
  Ishihara,~H. Optical trapping of quantum dots based on gap-mode-excitation of
  localized surface plasmon. \emph{The Journal of Physical Chemistry Letters}
  \textbf{2010}, \emph{1}, 2327--2333, doi: 10.1021/jz100659x\relax
\mciteBstWouldAddEndPuncttrue
\mciteSetBstMidEndSepPunct{\mcitedefaultmidpunct}
{\mcitedefaultendpunct}{\mcitedefaultseppunct}\relax
\EndOfBibitem
\bibitem[Jensen \latin{et~al.}(2016)Jensen, Huang, Chen, Choy, Bischof,
  Lončar, and Bawendi]{Bawendi2016Bowtie}
Jensen,~R.~A.; Huang,~I.~C.; Chen,~O.; Choy,~J.~T.; Bischof,~T.~S.;
  Lončar,~M.; Bawendi,~M.~G. Optical trapping and two-photon excitation of
  colloidal quantum dots using bowtie apertures. \emph{ACS Photonics}
  \textbf{2016}, \emph{3}, 423--427\relax
\mciteBstWouldAddEndPuncttrue
\mciteSetBstMidEndSepPunct{\mcitedefaultmidpunct}
{\mcitedefaultendpunct}{\mcitedefaultseppunct}\relax
\EndOfBibitem
\bibitem[Hong \latin{et~al.}(2021)Hong, Yang, Kravchenko, and
  Ndukaife]{Ndukaife2021ElecTherPlas}
Hong,~C.; Yang,~S.; Kravchenko,~I.~I.; Ndukaife,~J.~C. Electrothermoplasmonic
  trapping and dynamic manipulation of single colloidal nanodiamond. \emph{Nano
  Letters} \textbf{2021}, \emph{21}, 4921--4927, doi:
  10.1021/acs.nanolett.1c00357\relax
\mciteBstWouldAddEndPuncttrue
\mciteSetBstMidEndSepPunct{\mcitedefaultmidpunct}
{\mcitedefaultendpunct}{\mcitedefaultseppunct}\relax
\EndOfBibitem
\bibitem[Chen \latin{et~al.}(2012)Chen, Serey, Sarkar, Chen, and
  Erickson]{Erickson2012Photonic}
Chen,~Y.-F.; Serey,~X.; Sarkar,~R.; Chen,~P.; Erickson,~D. Controlled photonic
  manipulation of proteins and other nanomaterials. \emph{Nano Letters}
  \textbf{2012}, \emph{12}, 1633--1637, doi: 10.1021/nl204561r\relax
\mciteBstWouldAddEndPuncttrue
\mciteSetBstMidEndSepPunct{\mcitedefaultmidpunct}
{\mcitedefaultendpunct}{\mcitedefaultseppunct}\relax
\EndOfBibitem
\bibitem[Xu and Crozier(2019)Xu, and Crozier]{Xu2019AlldielectricNF}
Xu,~Z.; Crozier,~K.~B. All-dielectric nanotweezers for trapping and observation
  of a single quantum dot. \emph{Optics Express} \textbf{2019}, \emph{27},
  4034--4045\relax
\mciteBstWouldAddEndPuncttrue
\mciteSetBstMidEndSepPunct{\mcitedefaultmidpunct}
{\mcitedefaultendpunct}{\mcitedefaultseppunct}\relax
\EndOfBibitem
\bibitem[Bouloumis and {Nic Chormaic}(2020)Bouloumis, and {Nic
  Chormaic}]{Bouloumis2020FromFT}
Bouloumis,~T.~D.; {Nic Chormaic},~S. From far-field to near-field micro- and
  nanoparticle optical trapping. \emph{Applied Sciences} \textbf{2020},
  \emph{10}, 1375\relax
\mciteBstWouldAddEndPuncttrue
\mciteSetBstMidEndSepPunct{\mcitedefaultmidpunct}
{\mcitedefaultendpunct}{\mcitedefaultseppunct}\relax
\EndOfBibitem
\bibitem[Bouloumis \latin{et~al.}(2020)Bouloumis, Kotsifaki, Han, {Nic
  Chormaic}, and Truong]{Bouloumis2020FastAE}
Bouloumis,~T.; Kotsifaki,~D.~G.; Han,~X.; {Nic Chormaic},~S.; Truong,~V.~G.
  Fast and efficient nanoparticle trapping using plasmonic connected nanoring
  apertures. \emph{Nanotechnology} \textbf{2020}, \emph{32}, 025507\relax
\mciteBstWouldAddEndPuncttrue
\mciteSetBstMidEndSepPunct{\mcitedefaultmidpunct}
{\mcitedefaultendpunct}{\mcitedefaultseppunct}\relax
\EndOfBibitem
\bibitem[Zehtabi-Oskuie \latin{et~al.}(2013)Zehtabi-Oskuie, Jiang, Cyr,
  Rennehan, Al-Balushi, and Gordon]{Gordon2013Hopping}
Zehtabi-Oskuie,~A.; Jiang,~H.; Cyr,~B.~R.; Rennehan,~D.~W.; Al-Balushi,~A.~A.;
  Gordon,~R. Double nanohole optical trapping: Dynamics and protein-antibody
  co-trapping. \emph{Lab on a Chip} \textbf{2013}, \emph{13}, 2563--2568\relax
\mciteBstWouldAddEndPuncttrue
\mciteSetBstMidEndSepPunct{\mcitedefaultmidpunct}
{\mcitedefaultendpunct}{\mcitedefaultseppunct}\relax
\EndOfBibitem
\bibitem[Yoon \latin{et~al.}(2020)Yoon, Song, Lee, Kim, Lee, and
  Kim]{Yoon2020HoppingOS}
Yoon,~S.~J.; Song,~D.; Lee,~J.; Kim,~M.-K.; Lee,~Y.-H.; Kim,~C.-K. Hopping of
  single nanoparticles trapped in a plasmonic double-well potential.
  \emph{Nanophotonics} \textbf{2020}, \emph{9}, 4729 -- 4735\relax
\mciteBstWouldAddEndPuncttrue
\mciteSetBstMidEndSepPunct{\mcitedefaultmidpunct}
{\mcitedefaultendpunct}{\mcitedefaultseppunct}\relax
\EndOfBibitem
\bibitem[Park \latin{et~al.}(2008)Park, Lee, Galloway, and Searson]{CdQDs2008}
Park,~J.; Lee,~K.~H.; Galloway,~J.~F.; Searson,~P.~C. Synthesis of Cadmium
  Selenide Quantum Dots from a Non-Coordinating Solvent: Growth Kinetics and
  Particle Size Distribution. \emph{The Journal of Physical Chemistry C}
  \textbf{2008}, \emph{112}, 17849--17854\relax
\mciteBstWouldAddEndPuncttrue
\mciteSetBstMidEndSepPunct{\mcitedefaultmidpunct}
{\mcitedefaultendpunct}{\mcitedefaultseppunct}\relax
\EndOfBibitem
\bibitem[Shekhirev \latin{et~al.}(2017)Shekhirev, Goza, Teeter, Lipatov, and
  Sinitskii]{LeadQDs2017}
Shekhirev,~M.; Goza,~J.; Teeter,~J.~D.; Lipatov,~A.; Sinitskii,~A. Synthesis of
  Cesium Lead Halide Perovskite Quantum Dots. \emph{Journal of Chemical
  Education} \textbf{2017}, \emph{94}, 1150--1156\relax
\mciteBstWouldAddEndPuncttrue
\mciteSetBstMidEndSepPunct{\mcitedefaultmidpunct}
{\mcitedefaultendpunct}{\mcitedefaultseppunct}\relax
\EndOfBibitem
\bibitem[Tian \latin{et~al.}(2016)Tian, Zhong, Wu, Jiang, Shen, Jiang, and
  Wang]{TIAN2016204}
Tian,~R.; Zhong,~S.; Wu,~J.; Jiang,~W.; Shen,~Y.; Jiang,~W.; Wang,~T.
  Solvothermal method to prepare graphene quantum dots by hydrogen peroxide.
  \emph{Optical Materials} \textbf{2016}, \emph{60}, 204--208\relax
\mciteBstWouldAddEndPuncttrue
\mciteSetBstMidEndSepPunct{\mcitedefaultmidpunct}
{\mcitedefaultendpunct}{\mcitedefaultseppunct}\relax
\EndOfBibitem
\bibitem[Stenspil and Laursen(2024)Stenspil, and Laursen]{Photofluo}
Stenspil,~S.~G.; Laursen,~B.~W. Photophysics of fluorescent nanoparticles based
  on organic dyes – challenges and design principles. \emph{Chemical Science}
  \textbf{2024}, \emph{15}, 8625--8638\relax
\mciteBstWouldAddEndPuncttrue
\mciteSetBstMidEndSepPunct{\mcitedefaultmidpunct}
{\mcitedefaultendpunct}{\mcitedefaultseppunct}\relax
\EndOfBibitem
\bibitem[Paternò \latin{et~al.}(2021)Paternò, Goudappagouda, Chen, Lanzani,
  Scotognella, and Narita]{Narita2021Poly}
Paternò,~G.~M.; Goudappagouda,; Chen,~Q.; Lanzani,~G.; Scotognella,~F.;
  Narita,~A. Large Polycyclic Aromatic Hydrocarbons as Graphene Quantum Dots:
  from Synthesis to Spectroscopy and Photonics. \emph{Advanced Optical
  Materials} \textbf{2021}, \emph{9}, 2100508\relax
\mciteBstWouldAddEndPuncttrue
\mciteSetBstMidEndSepPunct{\mcitedefaultmidpunct}
{\mcitedefaultendpunct}{\mcitedefaultseppunct}\relax
\EndOfBibitem
\bibitem[Hu \latin{et~al.}(2017)Hu, Wang, Peng, Wang, Cao, Feng, Müllen, and
  Narita]{Narita2017D7H}
Hu,~Y.; Wang,~X.-Y.; Peng,~P.-X.; Wang,~X.-C.; Cao,~X.-Y.; Feng,~X.;
  Müllen,~K.; Narita,~A. Benzo-fused double [7]carbohelicene: Synthesis,
  structures, and physicochemical properties. \emph{Angewandte Chemie
  International Edition} \textbf{2017}, \emph{56}, 3374--3378\relax
\mciteBstWouldAddEndPuncttrue
\mciteSetBstMidEndSepPunct{\mcitedefaultmidpunct}
{\mcitedefaultendpunct}{\mcitedefaultseppunct}\relax
\EndOfBibitem
\bibitem[Kotsifaki \latin{et~al.}(2020)Kotsifaki, Truong, and {Nic
  Chormaic}]{Kotsifaki2020FanoResonantAM}
Kotsifaki,~D.~G.; Truong,~V.~G.; {Nic Chormaic},~S. Fano-resonant, asymmetric,
  metamaterial-assisted tweezers for single nanoparticle trapping. \emph{Nano
  Letters} \textbf{2020}, \emph{20}, 3388–3395\relax
\mciteBstWouldAddEndPuncttrue
\mciteSetBstMidEndSepPunct{\mcitedefaultmidpunct}
{\mcitedefaultendpunct}{\mcitedefaultseppunct}\relax
\EndOfBibitem
\bibitem[Bouloumis \latin{et~al.}(2023)Bouloumis, Kotsifaki, and {Nic
  Chormaic}]{BouloumisEnabling}
Bouloumis,~T.~D.; Kotsifaki,~D.~G.; {Nic Chormaic},~S. Enabling self-induced
  back-action trapping of gold nanoparticles in metamaterial plasmonic
  tweezers. \emph{Nano Letters} \textbf{2023}, \emph{23}, 4723--4731\relax
\mciteBstWouldAddEndPuncttrue
\mciteSetBstMidEndSepPunct{\mcitedefaultmidpunct}
{\mcitedefaultendpunct}{\mcitedefaultseppunct}\relax
\EndOfBibitem
\bibitem[Kotsifaki \latin{et~al.}(2023)Kotsifaki, Singh, {Nic Chormaic}, and
  Truong]{KotsifakiColi2023}
Kotsifaki,~D.~G.; Singh,~R.~R.; {Nic Chormaic},~S.; Truong,~V.~G. Asymmetric
  split-ring plasmonic nanostructures for the optical sensing of Escherichia
  coli. \emph{Biomedical Optics Express} \textbf{2023}, \emph{14},
  4875--4887\relax
\mciteBstWouldAddEndPuncttrue
\mciteSetBstMidEndSepPunct{\mcitedefaultmidpunct}
{\mcitedefaultendpunct}{\mcitedefaultseppunct}\relax
\EndOfBibitem
\bibitem[Kotsifaki \latin{et~al.}(2024)Kotsifaki, Truong, Dindo, Laurino, and
  {Nic Chormaic}]{kotsifaki2024hybrid}
Kotsifaki,~D.~G.; Truong,~V.~G.; Dindo,~M.; Laurino,~P.; {Nic Chormaic},~S.
  Hybrid Metamaterial Optical Tweezers for Dielectric Particles and
  Biomolecules Discrimination. \emph{arXiv} \textbf{2024},
  \emph{physics.optics}, 2402.12878\relax
\mciteBstWouldAddEndPuncttrue
\mciteSetBstMidEndSepPunct{\mcitedefaultmidpunct}
{\mcitedefaultendpunct}{\mcitedefaultseppunct}\relax
\EndOfBibitem
\bibitem[Jiang \latin{et~al.}(2020)Jiang, Rogez, Claude, Baffou, and
  Wenger]{Jiang2020QuantifyingTR}
Jiang,~Q.; Rogez,~B.; Claude,~J.-B.; Baffou,~G.; Wenger,~J. Quantifying the
  role of the surfactant and the thermophoretic force in plasmonic nano-optical
  trapping. \emph{Nano Letters} \textbf{2020}, \emph{20}, 8811–8817\relax
\mciteBstWouldAddEndPuncttrue
\mciteSetBstMidEndSepPunct{\mcitedefaultmidpunct}
{\mcitedefaultendpunct}{\mcitedefaultseppunct}\relax
\EndOfBibitem
\bibitem[Kotsifaki and {Nic Chormaic}(2022)Kotsifaki, and {Nic
  Chormaic}]{Kotsifaki2022Thermal}
Kotsifaki,~D.~G.; {Nic Chormaic},~S. The role of temperature-induced effects
  generated by plasmonic nanostructures on particle delivery and manipulation:
  a review. \emph{Nanophotonics} \textbf{2022}, \emph{11}, 2199--2218\relax
\mciteBstWouldAddEndPuncttrue
\mciteSetBstMidEndSepPunct{\mcitedefaultmidpunct}
{\mcitedefaultendpunct}{\mcitedefaultseppunct}\relax
\EndOfBibitem
\bibitem[Rodr{\'i}guez-Sevilla \latin{et~al.}(2018)Rodr{\'i}guez-Sevilla,
  Prorok, Bednarkiewicz, Marqu{\'e}s, Garc{\'i}a-Mart{\'i}n, Sol{\'e},
  Haro‐Gonz{\'a}lez, and Jaque]{RodrguezSevilla2018OpticalFA}
Rodr{\'i}guez-Sevilla,~P.; Prorok,~K.; Bednarkiewicz,~A.; Marqu{\'e}s,~M.~I.;
  Garc{\'i}a-Mart{\'i}n,~A.; Sol{\'e},~J. A.~G.; Haro‐Gonz{\'a}lez,~P.;
  Jaque,~D. Optical forces at the nanoscale: Size and electrostatic effects.
  \emph{Nano Letters} \textbf{2018}, \emph{18}, 602--609\relax
\mciteBstWouldAddEndPuncttrue
\mciteSetBstMidEndSepPunct{\mcitedefaultmidpunct}
{\mcitedefaultendpunct}{\mcitedefaultseppunct}\relax
\EndOfBibitem
\bibitem[Ko \latin{et~al.}(2007)Ko, Park, Kim, and Kim]{SuKo_Ga}
Ko,~D.-S.; Park,~Y.~M.; Kim,~S.-D.; Kim,~Y.-W. Effective removal of Ga residue
  from focused ion beam using a plasma cleaner. \emph{Ultramicroscopy}
  \textbf{2007}, \emph{107}, 368--373\relax
\mciteBstWouldAddEndPuncttrue
\mciteSetBstMidEndSepPunct{\mcitedefaultmidpunct}
{\mcitedefaultendpunct}{\mcitedefaultseppunct}\relax
\EndOfBibitem
\bibitem[Bouloumis(2023)]{TheoThesis}
Bouloumis,~T. Metamaterial plasmonic tweezers for enhanced nanoparticle
  trapping, Ch. 3 \& 4. Ph.D.\ thesis, Okinawa Institute of Science and
  Technology Graduate University, 2023;
  \url{https://doi.org/10.15102/1394.00002679}\relax
\mciteBstWouldAddEndPuncttrue
\mciteSetBstMidEndSepPunct{\mcitedefaultmidpunct}
{\mcitedefaultendpunct}{\mcitedefaultseppunct}\relax
\EndOfBibitem
\bibitem[Raki\'{c} \latin{et~al.}(1998)Raki\'{c}, Djuri\v{s}i\'{c}, Elazar, and
  Majewski]{Rakic:98}
Raki\'{c},~A.~D.; Djuri\v{s}i\'{c},~A.~B.; Elazar,~J.~M.; Majewski,~M.~L.
  Optical properties of metallic films for vertical-cavity optoelectronic
  devices. \emph{Applied Optics} \textbf{1998}, \emph{37}, 5271--5283\relax
\mciteBstWouldAddEndPuncttrue
\mciteSetBstMidEndSepPunct{\mcitedefaultmidpunct}
{\mcitedefaultendpunct}{\mcitedefaultseppunct}\relax
\EndOfBibitem
\bibitem[Kotsifaki and {Nic Chormaic}(2019)Kotsifaki, and {Nic
  Chormaic}]{Kotsifaki2019PlasmonicOT}
Kotsifaki,~D.~G.; {Nic Chormaic},~S. Plasmonic optical tweezers based on
  nanostructures: fundamentals, advances and prospects. \emph{Nanophotonics}
  \textbf{2019}, \emph{8}, 1227 -- 1245\relax
\mciteBstWouldAddEndPuncttrue
\mciteSetBstMidEndSepPunct{\mcitedefaultmidpunct}
{\mcitedefaultendpunct}{\mcitedefaultseppunct}\relax
\EndOfBibitem
\bibitem[Roxworthy \latin{et~al.}(2014)Roxworthy, Bhuiya, Vanka, and
  Toussaint]{Roxworthy2014Convection}
Roxworthy,~B.~J.; Bhuiya,~A.~M.; Vanka,~S.~P.; Toussaint,~K.~C. Understanding
  and controlling plasmon-induced convection. \emph{Nature Communications}
  \textbf{2014}, \emph{5}, 3173\relax
\mciteBstWouldAddEndPuncttrue
\mciteSetBstMidEndSepPunct{\mcitedefaultmidpunct}
{\mcitedefaultendpunct}{\mcitedefaultseppunct}\relax
\EndOfBibitem
\bibitem[Braibanti \latin{et~al.}(2008)Braibanti, Vigolo, and
  Piazza]{Braibanti}
Braibanti,~M.; Vigolo,~D.; Piazza,~R. Does thermophoretic mobility depend on
  particle size? \emph{Physical Review Letters} \textbf{2008}, \emph{100},
  108303\relax
\mciteBstWouldAddEndPuncttrue
\mciteSetBstMidEndSepPunct{\mcitedefaultmidpunct}
{\mcitedefaultendpunct}{\mcitedefaultseppunct}\relax
\EndOfBibitem
\bibitem[Baffou \latin{et~al.}(2013)Baffou, Berto, Bermúdez~Ureña, Quidant,
  Monneret, Polleux, and Rigneault]{Baffou2013ThermoArrays}
Baffou,~G.; Berto,~P.; Bermúdez~Ureña,~E.; Quidant,~R.; Monneret,~S.;
  Polleux,~J.; Rigneault,~H. Photoinduced heating of nanoparticle arrays.
  \emph{ACS Nano} \textbf{2013}, \emph{7}, 6478--6488\relax
\mciteBstWouldAddEndPuncttrue
\mciteSetBstMidEndSepPunct{\mcitedefaultmidpunct}
{\mcitedefaultendpunct}{\mcitedefaultseppunct}\relax
\EndOfBibitem
\bibitem[Fränzl and Cichos(2022)Fränzl, and Cichos]{Cichos2022}
Fränzl,~M.; Cichos,~F. Hydrodynamic manipulation of nano-objects by optically
  induced thermo-osmotic flows. \emph{Nature Communications} \textbf{2022},
  \emph{13}, 656\relax
\mciteBstWouldAddEndPuncttrue
\mciteSetBstMidEndSepPunct{\mcitedefaultmidpunct}
{\mcitedefaultendpunct}{\mcitedefaultseppunct}\relax
\EndOfBibitem
\bibitem[Jiang \latin{et~al.}(2019)Jiang, Rogez, Claude, Baffou, and
  Wenger]{Wenger2019TemperatureMeasur}
Jiang,~Q.; Rogez,~B.; Claude,~J.-B.; Baffou,~G.; Wenger,~J. Temperature
  measurement in plasmonic nanoapertures used for optical trapping. \emph{ACS
  Photonics} \textbf{2019}, \emph{6}, 1763--1773\relax
\mciteBstWouldAddEndPuncttrue
\mciteSetBstMidEndSepPunct{\mcitedefaultmidpunct}
{\mcitedefaultendpunct}{\mcitedefaultseppunct}\relax
\EndOfBibitem
\bibitem[Würger(2010)]{WurgerSoret}
Würger,~A. Thermal non-equilibrium transport in colloids. \emph{Reports on
  Progress in Physics} \textbf{2010}, \emph{73}, 126601\relax
\mciteBstWouldAddEndPuncttrue
\mciteSetBstMidEndSepPunct{\mcitedefaultmidpunct}
{\mcitedefaultendpunct}{\mcitedefaultseppunct}\relax
\EndOfBibitem
\bibitem[Kotnala \latin{et~al.}(2020)Kotnala, Kollipara, Li, and
  Zheng]{Kotnala2019Overcomingdiffusion}
Kotnala,~A.; Kollipara,~P.~S.; Li,~J.; Zheng,~Y. Overcoming diffusion-limited
  trapping in nanoaperture tweezers using opto-thermal-induced flow. \emph{Nano
  Letters} \textbf{2020}, \emph{20}, 768--779, doi:
  10.1021/acs.nanolett.9b04876\relax
\mciteBstWouldAddEndPuncttrue
\mciteSetBstMidEndSepPunct{\mcitedefaultmidpunct}
{\mcitedefaultendpunct}{\mcitedefaultseppunct}\relax
\EndOfBibitem
\bibitem[Jones \latin{et~al.}(2015)Jones, Maragò, and
  Volpe]{jones_marago_volpe_2015}
Jones,~P.~H.; Maragò,~O.~M.; Volpe,~G. \emph{Optical Tweezers: Principles and
  Applications}; Cambridge University Press, 2015\relax
\mciteBstWouldAddEndPuncttrue
\mciteSetBstMidEndSepPunct{\mcitedefaultmidpunct}
{\mcitedefaultendpunct}{\mcitedefaultseppunct}\relax
\EndOfBibitem
\bibitem[Mestres \latin{et~al.}(2016)Mestres, Berthelot, Acimovic, and
  Quidant]{Mestres2016Unravelling}
Mestres,~P.; Berthelot,~J.; Acimovic,~S.~S.; Quidant,~R. Unraveling the
  optomechanical nature of plasmonic trapping. \emph{Light: Science \&
  Applications} \textbf{2016}, \emph{5}, e16092\relax
\mciteBstWouldAddEndPuncttrue
\mciteSetBstMidEndSepPunct{\mcitedefaultmidpunct}
{\mcitedefaultendpunct}{\mcitedefaultseppunct}\relax
\EndOfBibitem
\end{mcitethebibliography}

\end{document}